\documentclass[journal,draftclsnofoot,onecolumn]{IEEEtran}
\usepackage{amsmath}
\usepackage{amssymb}
\usepackage{amsthm}
\usepackage{color}
\usepackage{url}
\usepackage{comment}
\usepackage{eucal}

\usepackage{algorithm}
\usepackage{algpseudocode}

\usepackage{mathtools}
\DeclarePairedDelimiter{\set}{\{}{\}}

\usepackage{amsfonts,bm}
\renewcommand{\mathbf}[1]{{\bm{#1}}}
\renewenvironment{proof}{\begin{IEEEproof}}{\end{IEEEproof}}

\DeclarePairedDelimiter\abs{\lvert}{\rvert}
\DeclarePairedDelimiter\norm{\lVert}{\rVert}
\DeclarePairedDelimiter\ceil{\lceil}{\rceil}
\DeclarePairedDelimiter\floor{\lfloor}{\rfloor}
\DeclarePairedDelimiter\parenv{\lparen}{\rparen}

\newcommand{\cJ}{{\mathcal{J}}}
\newcommand{\cK}{{\mathcal{K}}}
\newcommand{\cP}{{\mathcal{P}}}
\newcommand{\cR}{{\mathcal{R}}}
\newcommand{\cT}{{\mathcal{T}}}

\renewcommand{\le}{\leqslant}
\renewcommand{\ge}{\geqslant}

\theoremstyle{plain}
\newtheorem{theorem}{\indent Theorem}
\newtheorem{corollary}[theorem]{\indent Corollary}
\newtheorem{lemma}[theorem]{\indent Lemma}
\newtheorem{proposition}[theorem]{\indent Proposition}

\theoremstyle{definition}

\newtheorem{example}{\indent Example}
\theoremstyle{remark}
\newtheorem{remark}{\indent Remark}

\newcommand{\F}{\mathbb{F}}
\newcommand{\R}{\mathbb{R}}
\newcommand{\Z}{\mathbb{Z}}

\newcommand{\Zero}{{\mathbf{0}}}
\newcommand{\poly}{{\mathrm{poly}}}

\newcommand{\Sinn}{{\mathcal{S}}}
\newcommand{\eqdef}{\triangleq}
\newcommand{\TD}{{\mathrm{TD}}}
\newcommand{\Block}{{\mathfrak{B}}}
\newcommand{\Group}{{\mathfrak{G}}}

\newcommand{\bldc}{{\mathbf{c}}}
\newcommand{\blde}{{\mathbf{e}}}
\newcommand{\bldh}{{\mathbf{h}}}
\newcommand{\blds}{{\mathbf{s}}}
\newcommand{\bldu}{{\mathbf{u}}}
\newcommand{\bldx}{{\mathbf{x}}}
\newcommand{\bldy}{{\mathbf{y}}}

\newcommand{\bldchi}{{\mathbf{\chi}}}
\newcommand{\bldepsilon}{\pmb{\varepsilon}}
\newcommand{\code}{{\mathcal{C}}}
\newcommand{\Decoder}{{\mathcal{D}}}
\newcommand{\Decodert}{\widetilde{\Decoder}}
\newcommand{\Decoderh}{\widehat{\Decoder}}
\newcommand{\Decoderb}{\overline{\Decoder}}
\newcommand{\Support}{{\mathsf{Supp}}}
\newcommand{\weight}{{\mathsf{w}}}
\newcommand{\distance}{{\mathsf{d}}}
\newcommand{\Height}{{\mathsf{h}}}
\newcommand{\Ball}{{\mathcal{B}}}
\newcommand{\Cube}{{\mathcal{Q}}}
\newcommand{\transpose}{{\top\scriptscriptstyle{\!}}}
\newcommand{\hp}{{\lambda}}
\newcommand{\an}{{a}} 
\newcommand{\failure}{{\mathrm{``e"}}}
\newcommand{\zeroone}{{0\textrm{--}1}}
\newcommand{\Int}[1]{{\left[{#1}\right]}}
\renewcommand{\qed}{\endIEEEproof}

\begin{document}
\date{}
\title{Multiple-Error-Correcting Codes for \\
Analog Computing on Resistive Crossbars}

\author{%
Hengjia Wei and Ron M. Roth%
\thanks{Hengjia Wei is with the Peng Cheng Laboratory, Shenzhen 518055,
China (e-mail: hjwei05@gmail.com).
He is also with the School of Mathematics and Statistics,
Xi'an Jiaotong University, Xi'an 710049, China, and
the Pazhou Laboratory (Huangpu), Guangzhou 510555, China.}%
\thanks{Ron M. Roth is with the Computer Science Department,
Technion--–Israel Institute of Technology, Haifa 3200003, Israel
(e-mail: ronny@cs.technion.ac.il).}
\thanks{The work of H.~Wei was supported in part by the major key
project of Peng Cheng Laboratory under grant PCL2023AS1-2 and
the National Natural Science Foundation of China under Grant 12371523.
The work of R.~M.~Roth was supported in part by
Grant No.~1713/20 from the Israel Science Foundation.}%
}

\maketitle

\begin{abstract}
Error-correcting codes over the real field are studied which can locate
outlying computational errors
when performing approximate computing of real vector--matrix
multiplication on resistive crossbars.
Prior work has concentrated on locating a single outlying error and,
in this work, several classes of codes are presented which can handle
multiple errors. It is first shown that one of the known constructions,
which is based on spherical codes, can in fact handle multiple
outlying errors. A second family of codes is then presented
with $\zeroone$~parity-check matrices which are sparse and disjunct;
such matrices have been used in other applications as well,
especially in combinatorial group testing.
In addition, a certain class of the codes that are obtained through
this construction is shown to be efficiently decodable.
As part of the study of sparse disjunct matrices,
this work also contains improved lower and upper bounds on
the maximum Hamming weight of the rows in such matrices.
\end{abstract}


\section{Introduction}
\label{sec:introduction}

Vector--matrix multiplication is a computational task
that is found in numerous applications,
including machine learning (e.g., deep learning)
and signal processing. Designing circuits for
vector--matrix multiplication requires achieving high computational
throughput while concurrently ensuring minimal energy consumption and
a compact physical footprint. These criteria have prompted recent
proposals to incorporate resistive memory technology into analog
computing architectures.

Let~$\bldu$ be a row $\ell$-vector and~$A$ be
an $\ell \times n$ matrix---both with (nonnegative)
entries in~$\R$ or~$\Z$.
In current implementations of
vector--matrix multiplication~%
\cite{Boseretal91},%
\cite{Huetal18},%
\cite{Huetal16},%
\cite{Kubetal90},%
\cite{Shafetal16},
the matrix $A = (a_{i,j})$ is realized as a crossbar of
$\ell$~row conductors and $n$ column conductors with
programmable nano-scale resistors at the junctions.
The resistor at the junction $(i,j)$ is set to have conductance that is
proportional to the entry $a_{i,j}$ of~$A$.
Each entry $u_i$ of~$\bldu$ is converted into a voltage level that is
proportional to $u_i$ and fed to the corresponding row conductor.
Then the product $\bldc = \bldu A$, carried out over
the real field $\R$, can be computed by reading the currents
at the column conductors.
Negative entries in~$\bldu$ or~$A$ can be accommodated
by duplication of the circuit.

Recently, the second author proposed two classes of coding schemes
to locate computational errors under two distinct scenarios:
\emph{exact} integer vector--matrix multiplication~\cite{Roth19}
and \emph{approximate} real vector--matrix multiplication~\cite{Roth20}.
We next describe the second scenario, as it will be
the subject of this work as well.

In the model described in~\cite{Roth20}, the ideal computation
$\bldc =\bldu A \in \R^n$ may be distorted
by two types of errors, which lead to a read vector
\begin{equation}
\label{eq:errmodel}
\bldy = \bldc + \bldepsilon+\blde \in \R^n,
\end{equation}
where $\blde, \bldepsilon \in \R^n$.
The entries of~$\bldepsilon$ are all within the interval
$[-\delta,\delta]$ for some prescribed threshold~$\delta$, representing
small computational errors that are tolerable,
while the entries of~$\blde$ represent outlying errors that may be
caused by events such as stuck cells or short cells in the array
(and may have large magnitudes).
The goal is to design a coding scheme that allows to locate all
the non-zero entries of~$\blde$ that are outside
an interval $[-\Delta,\Delta]$, for the smallest~$\Delta$, provided that
the number of outlying errors does not exceed
a prescribed number~$\tau$.
A more general setting includes the option of detecting
$\sigma$ additional errors and, as shown in~\cite{Roth20},
in this case the value $\hp = 2\tau + \sigma$ plays a role
when analyzing the correction capability of a coding scheme.

The encoding scheme presented in~\cite{Roth20} can be characterized
by a linear $[n,k]$ code~$\code$ over~$\R$:
we allocate $r = n-k$ columns of the matrix~$A$ for redundancy so that
each row of~$A$ forms a codeword of~$\code$. Then the result of
the multiplication of any input real row vector~$\bldu$ by
the matrix~$A$ is also a codeword of~$\code$.

In crude terms
(with more details to be provided in Section~\ref{sec:pre}),
the required condition from the linear code~$\code$ is that
it has a decoder that locates all the outlying errors of magnitude
above~$\Delta$, whenever the Hamming weight
of~$\blde$ does not exceed~$\tau$; moreover, if the decoder
returns a set of locations (rather than just detects errors),
then~$\blde$ should be nonzero at all these locations.
Linear codes over~$\R$ which satisfy this condition are referred
to as \emph{analog error-correcting codes}.

For the case $\hp = 2\tau + \sigma \le 2$
(which includes the single error location/detection cases,
i.e., $(\tau,\sigma) = (0,1), (1,0)$),
code constructions were proposed in~\cite{Roth20}
for several trade-offs between
the redundancy~$r$ and the smallest attainable ratio $\Delta/\delta$.
One of the constructions for $\hp = 2$ has a sparse parity-check matrix
over $\set{-1, 0, 1}$ and attains
$\Delta/\delta \le 2 \ceil{2n/r}$,
for every even redundancy $r \ge \sqrt{n}$;
another construction has a parity-check matrix that forms
a spherical code and attains $\Delta/\delta = O(n/\sqrt{r})$
with $r = \Theta(\log n)$.

In this work, we present several classes of codes over~$\R$
for a wide range of values~$\hp$,
and compute upper bounds on the attainable ratios
$\Delta/\delta$, in terms of $n$, $r$, and~$\hp$;
see Table~\ref{tab:summary}.
When $\hp = 2$, our bounds coincide with
those presented in~\cite{Roth20},\cite{Roth2022ITW}.
One of the classes is actually the spherical code scheme
of~\cite{Roth20} when constructed with
redundancy $r = \Theta(\hp^2 \log n)$:
we show that these codes can still attain
$\Delta/\delta = O(n/\sqrt{r})$ yet for a wide range of $\hp \ge 2$.
In our analysis we make use of
the \emph{restricted isometry property}
(and a variant thereof)
of \emph{matrices of low coherence}---a tool which is widely used
in compressed sensing~\cite{Bouretal},\cite{Candes2008}.

A second class of codes to be presented is based on
\emph{disjunct matrices with limited row weights}---a notion that
has been applied, \emph{inter alia},
in combinatorial group testing~\cite{InaKaiOzg20}.
Employing the known construction of disjunct matrices
of~\cite{InaKaiOzg20}, for any $n, \ell, \hp \in \Z^+$ such that
$n^{1/(\ell+1)}$ is a prime power
and $\hp \le \ceil{n^{1/(\ell+1)}/\ell}$,
our codes attain $\Delta/\delta \le 2n^{\ell/(\ell+1)}$
with $r \le \ell \hp n^{1/(\ell+1)}$.

Our study also includes a new family of disjunct matrices
(which, in turn, can then be employed in our code construction
mentioned above).
Specifically, for any positive integer $\rho \le \sqrt{n}$
such that $n/\rho$ is a prime power,
we construct optimal disjunct matrices
with maximum row weight${} \le \rho$,
achieving the lower bound on the number of rows
as stated in~\cite{InaKaiOzg20}; formerly, such disjunct matrices
were exclusively established for $\rho = \sqrt{n}$.
Moreover, by deriving a new lower bound on the number of rows,
we show that the construction in~\cite{InaKaiOzg20}
of disjunct matrices with maximum row weight
$\rho \ge \sqrt{n}$ is asymptotically optimal.

The paper is organized as follows. We begin, in Section~\ref{sec:pre},
by providing notation and known results used throughout the paper.
This section also contains our new lower bound on the number of rows
of a disjunct matrix with limited row weights.

In Section~\ref{sec:charmulerr}, we analyze the spherical code
construction and establish its performance for a wide range
of values~$\hp$.

In Section~\ref{sec:codefromdismat}, we present the code construction
that is based on disjunct matrices with limited row weights.
Efficient decoding algorithms to locate the outlying errors are also
presented.

In Section~\ref{sec:conofdismat}, we present our new family
of optimal disjunct matrices.
Employing these matrices in our code construction,
for any (fixed) rational number $\alpha \in [1/2,1)$
they attain $r \le \hp n^\alpha$ and $\Delta/\delta \le 2\hp n/r$
for infinitely many values of~$n$.
Interestingly, these parameters align with those of
the single-error-correcting codes in~\cite{Roth20}
(wherein $r$ can be any even integer such that $r(r-1) \ge n$
and $\Delta/\delta \le 2 \ceil{2n/r}$).

\begin{table*}
  \caption{Summary of the $[n,k{\ge}n{-}r]$ codes~$\code$ over~$\R$}
  \label{tab:summary}
\center
  {\renewcommand{\arraystretch}{2.5}
   \everymath={\displaystyle}
  \begin{tabular}{ccclll}
    \hline\hline
    $r$ & Attainable $\Delta/\delta$        &   $\hp$  &
    \multicolumn{1}{c}{Comments}  & Reference\\
    \hline
      $\Theta(\log n)$
    & $O\parenv*{\frac{n}{\sqrt{r}}}$  &    $2$
    &             &  Prop.~5 in~\cite{Roth2022ITW}       \\
    $\Theta(\hp^2 \log n)$
    & $O\parenv*{\frac{n}{\sqrt{r}}}$
    & $O\parenv*{\sqrt{\frac{n}{\log n}}}$
    &            &   Cor.~\ref{cor:newasymbound}       \\
    \hline
      $r\le n\le r(r-1)$
    & $ 2\ceil*{\frac{2n}{r}}$   &     $2$
    &            &     Prop.~6 in~\cite{Roth20}    \\
      $\frac{\hp n}{\rho}$
    & $\frac{2 \hp n}{r}$
    & $\hp \le \rho$
    & $\rho \in \Z^+$, $\rho \le \sqrt{n}$,
      $\frac{n}{\rho}$ is a prime power
    & Cor.~\ref{cor:code-2}\\
      $\frac{\hp n}{\rho}$
    & $\frac{2 \hp n}{r}$
    & $\hp \le \min \set{\rho,p_1^{e_1},p_2^{e_2},\ldots}$
    & $\rho \in \Z^+$, $\rho \le \sqrt{n}$,
      $\frac{n}{\rho} = p_1^{e_1} p_2^{e_2} \cdots$
    & Thm.~\ref{thm:TDdisj-2} \\
      $(\ell \hp - \ell + 1) q$
    & $\frac{2 (\ell\hp - \ell + 1) n}{r}$
    & $\hp \le \ceil{q/\ell}$
    & $\ell \in \Z^+$, $q$ is a prime power, $n = q^{\ell+1}$
    & Cor.~\ref{cor:code-1}\\
    \hline\hline
  \end{tabular}
  }
\end{table*}

\section{Preliminaries}
\label{sec:pre}

For integers $\ell \le n$, we denote by $[\ell:n]$ the integer subset
$\set*{z \in \Z \,:\, \ell \le z < n}$.
We will use the shorthand notation $\Int{n}$ for $\Int{0:n}$,
and we will typically use
$\Int{n}$ to index the entries of vectors in $\R^n$.
Similarly, the entries of an $r \times n$ matrix $H = (H_{i,j})$ will be
indexed by $(i,j) \in \Int{r} \times \Int{n}$, and $H_i$ and $\bldh_j$
will denote, respectively, row~$i$ and column~$j$ in~$H$.
For a subset $\cJ \subseteq \Int{n}$, the notation $(H)_\cJ$
stands for the $r \times \abs{\cJ}$ submatrix of~$H$
that is formed by the columns that are indexed by~$\cJ$.

Unless specified otherwise, all logarithms are taken to base~$2$.

\subsection{Analog error-correcting codes}
\label{sec:analogecc}

Given $\delta,\Delta \in \R^+$, let
\[
\Cube(n,\delta)
\eqdef \set*{\bldepsilon = (\varepsilon_j) \in \R^n \,:\,
\norm{\bldepsilon}_\infty \le \delta}
\]
be the set of all tolerable error vectors with threshold~$\delta$,
where $\norm{\bldepsilon}_\infty$ stands for
the infinity norm $\max_{j \in \Int{n}} \abs{\varepsilon_j}$.
For $\blde = (e_j)_j \in \R^n$, define
\[
\Support_\Delta(\blde)
\eqdef \set*{ j\in \Int{n} \,:\, \abs{e_j} > \Delta}.
\]
In particular, $\Support_0(\blde)$ is the ordinary support of~$\blde$.
We use $\weight(\blde)$ to denote the Hamming weight of~$\blde$.
The set of all vectors of Hamming weight at most~$w$
in $\R^n$ is denoted by $\Ball(n,w)$.

Let~$\code$ be a linear $[n,k]$ code over~$\R$.
A decoder for~$\code$ is a function
$\Decoder: \R^n \rightarrow 2^{\Int{n}} \cup \set{\failure}$
which returns a set of locations of outlying errors or
an indication~$\failure$ that errors have been detected.
Given $\delta, \Delta\in \R^+$ and prescribed nonnegative integers
$\tau$ and~$\sigma$, we say that the decoder~$\Decoder$
corrects $\tau$~errors and detects $\sigma$~additional errors
with respect to the threshold pair $(\delta,\Delta)$,
or that~$\Decoder$
is a \emph{$(\tau,\sigma)$-decoder for $(\code, \Delta:\delta)$},
if the following conditions hold for every~$\bldy$ as
in~\eqref{eq:errmodel}, where $\bldc\in \code$,
$\bldepsilon\in \Cube(n,\delta)$, and $\blde\in \Ball(n,\tau+\sigma)$.
\begin{description}
\item[(D1)]
If $\blde \in \Ball(n,\tau)$ then
$\failure \ne \Decoder(\bldy) \subseteq \Support_0(\blde)$.
\item[(D2)]
If $\Decoder(\bldy)\ne \failure$ then
$\Support_\Delta(\blde) \subseteq \Decoder(\bldy)$.
\end{description}

Let $\bldx = (x_j)_{j \in \Int{n}}$ be a nonzero vector in $\R^n$
and let~$\pi$ be a permutation on $\Int{n}$ such that
\[
\abs{x_{\pi(0)}} \ge \abs{x_{\pi(1)}}
\ge \cdots \ge \abs{x_{\pi(n-1)}}.
\]
Given an integer $\hp \in \Int{n}$,
the \emph{$\hp$-height} of~$\bldx$, denoted by $\Height_\hp(\bldx)$,
is defined as
\[
\Height_\hp(\bldx)
\eqdef \abs*{\frac{x_{\pi(0)}}{x_{\pi(\hp)}}} ,
\]
and we formally define $\Height_n(\bldx) \eqdef \infty$.
For a linear code $\code \ne \set{\Zero}$ over~$\R$, its $\hp$-height,
denoted by $\Height_\hp(\code)$, is defined by
\[
\Height_\hp(\code)
\eqdef \max_{\bldc \in \code\setminus\set{\Zero}} \Height_\hp(\bldc).
\]
The minimum Hamming distance of~$\code$, denoted by $\distance(\code)$,
can be related to $(\Height_\hp(\code))_\hp$ by
\begin{equation}
\label{eq:heighttomindist}
\distance(\code)
= \min \set{\hp \in \Int{n+1} \,:\, \Height_\hp(\code) = \infty}.
\end{equation}

\begin{theorem}[\cite{Roth20},\cite{Roth23}]
\label{thm:chaviamheight}
Let~$\code$ be a linear $[n,k]$ code over~$\R$.
There is a $(\tau,\sigma)$-decoder for $(\code,\Delta : \delta)$,
if and only if
\[
\Delta / \delta \ge 2 \, \Height_{2\tau+\sigma}(\code) + 2 .
\]
\end{theorem}

Theorem~\ref{thm:chaviamheight}
motivated in~\cite{Roth20} to define
for every $\hp \in \Int{n+1}$ the expression
\begin{equation}
\label{eq:defGamma}
\Gamma_\hp(\code) \eqdef 2 \, \Height_\hp(\code) + 2 ,
\end{equation}
so that $\Gamma_{2\tau+\sigma}(\code)$
is the smallest ratio $\Delta/\delta$ for which
there is a $(\tau,\sigma)$-decoder for $(\code,\Delta : \delta)$.
Equivalently,
$\Gamma_{2\tau+\sigma}$ is the smallest~$\Delta$ such that
there is a $(\tau,\sigma)$-decoder for $(\code,\Delta : 1)$.
Thus, given~$n$ and~$\hp$, our aim is to construct
linear codes~$\code$ over~$\R$ with both
$\Gamma_\hp(\code)$ and redundancy~$r$ as small as possible.

For the case $\hp = 2\tau + \sigma \le 2$,
a characterization of $\Gamma_1(\code)$ and $\Gamma_2(\code)$
was presented in~\cite{Roth20} in terms of
the parity-check matrix of~$\code$.
In the next proposition,
we present a generalization of that characterization
to any $\hp \in \Int{1:\distance(\code)}$.
Given a parity-check matrix~$H$ of~$\code$ over~$\R$, let
\begin{equation}
\label{eq:sinn}
\Sinn = \Sinn(H)
\eqdef
\set*{H \bldepsilon^\transpose \,:\, \bldepsilon \in \Cube(n,1)}
\end{equation}
and
\begin{eqnarray}
\label{eq:2sinn}
2 \Sinn
\eqdef
\Sinn + \Sinn
& = &
\set*{H (\bldepsilon + \bldepsilon')^\transpose \,:\,
\bldepsilon, \bldepsilon' \in \Cube(n,1)} \\
& = &
\set*{H \bldepsilon^\transpose \,:\, \bldepsilon \in \Cube(n,2)} .
\nonumber
\end{eqnarray}
Note that~$\Sinn$ is the set of
all the syndrome vectors (with respect to~$H$)
that can be obtained when there are no outlying errors,
assuming that $\delta = 1$.
Also, for $\Delta \in \R^+$ let
\begin{equation}
\label{eq:BsubDelta}
\Ball_\Delta(n,\hp)
\eqdef
\set*{\blde \in \Ball(n,\hp) \,:\, \norm{\blde}_\infty > \Delta} ,
\end{equation}
i.e., $\Ball_\Delta(n,\hp)$
consists of all the vectors $\blde \in \R^n$
such that both $\weight(\blde) \le \hp$
and $\Support_\Delta(\blde) \ne \varnothing$.

\begin{proposition}
\label{prop:alcharmultcrt}
Given a linear $[n,k{>}0]$ code~$\code$ over~$\R$,
let~$H$ be a parity-check matrix of~$\code$
and let $\hp \in \Int{1:\distance(\code)}$.
Then
\begin{equation}
\label{eq:alcharmultcrt}
\Gamma_\hp(\code) = \min \bigl\{ \Delta \in \R^+ :\,
H \blde^\transpose \notin 2 \Sinn
\;\; \textrm{for all $\blde \in \Ball_\Delta(n,\hp)$}
\bigr\} .
\end{equation}
\end{proposition}

\begin{proof}
We first show that $\Delta^* \eqdef \Gamma_\hp(\code)$ is contained
in the minimand set in~\eqref{eq:alcharmultcrt}.
Assume to the contrary
that there is a vector $\blde\in \Ball_{\Delta^*}(n,\hp)$
such that $H \blde^\transpose \in 2 \Sinn$, namely,
$H \blde^\transpose = H \bldepsilon^\transpose$
for some $\bldepsilon \in \Cube(n,2)$.
Then $\blde - \bldepsilon \in \code$ and, so,
\[
\Height_\hp(\code)\ge \Height_\hp (\blde-\bldepsilon)
\ge \frac{\norm{\blde}_\infty-2}{2} > \frac{\Delta^*-2}{2} .
\]
This, in turn, implies
\[
\Gamma_\hp(\code)
\stackrel{\textrm{\scriptsize \eqref{eq:defGamma}}}{=}
2 \, \Height_\hp(\code) + 2 > \Delta^* ,
\]
which is a contradiction.

We next show that $\Gamma_\hp(\code)$
is indeed the minimum of the set in~\eqref{eq:alcharmultcrt}.
Assuming to the contrary that this set contains
some $\Delta < \Gamma_\hp(\code)$,
there is a nonzero codeword $\bldc \in \code$ such that
\[
\Height_\hp(\bldc) > \frac{\Delta-2}{2}.
\]
Without loss of generality we can assume that
\[
c_0 \ge \abs{c_1} \ge \abs{c_2} \ge \cdots \ge \abs{c_{n-1}} ,
\]
where $\abs{c_\hp} = 2$ and (thus) $c_0 > \Delta - 2$.
Define the vectors $\blde, \bldepsilon \in \R^n$ as follows:
\[
\arraycolsep0.6ex
\begin{array}{ccc@{\!\;}ccccccccc@{\!\;}cc}
\blde       & = & ( & c_0{+}2 & c_1 & c_2 & \ldots & c_{\hp-1}
& 0     & 0         & \ldots & 0       & ) & , \\
\bldepsilon & = & ( & -2      & 0   & 0   & \ldots & 0
& c_\hp & c_{\hp+1} & \ldots & c_{n-1} & ) & .
\end{array}
\]
Then $\bldc = \blde + \bldepsilon$ and
$\bldepsilon \in \Cube(n,2)$, namely,
$H\blde^\transpose = -H\bldepsilon^\transpose \in 2 \Sinn$.
On the other hand $\blde \in \Ball_\Delta(n,\hp)$,
which means that~$\Delta$ is \emph{not}
in the minimand in~\eqref{eq:alcharmultcrt},
thereby reaching a contradiction.
\end{proof}

Propositions~9 and~8 in~\cite{Roth20} are
special cases of Proposition~\ref{prop:alcharmultcrt}
for $\hp = 1$ and $\hp = 2$, respectively.
Proposition~\ref{prop:alcharmultcrt} holds (vacuously)
also when $\hp \ge \distance(\code)$: in this case
the minimand in~\eqref{eq:alcharmultcrt} is empty
(since~$\code$ contains nonzero codewords in $\Ball(n,\hp)$
with arbitrary infinity norms),
while $\Gamma_\hp(\code) = \infty$ (from~\eqref{eq:heighttomindist}).

We end this subsection by mentioning two of the constructions
for $\hp = 2$ that were presented in~\cite{Roth20}.

\begin{theorem}[{\cite[Proposition~6]{Roth20}}]
\label{thm:rothcode}
Let~$H$ be an $r \times n$ matrix over $\set{-1,0,1}$
which satisfies the following three conditions:
\begin{enumerate}
\item
all columns of~$H$ are distinct,
\item
each column in~$H$ contains exactly two nonzero entries,
the first of which being a~$1$, and
\item
each row has Hamming weight $\floor{2n/r}$ or $\ceil{2n/r}$.
\end{enumerate}
(In particular, these conditions require that $n \le r(r-1)$.)
The linear $[n,k{\ge}n{-}r]$ code~$\code$ over~$\R$ with
a parity-check matrix~$H$ satisfies
$\Gamma_2(\code) \le 2\cdot \ceil{2n/r}$.
\end{theorem}

When~$r$ is even, the inequality $n \le r(r{-}1)$ is
also sufficient for having a matrix~$H$ that satisfies
the conditions of the theorem~\cite{KatonaSeress93}.

A second construction is presented in~\cite{Roth20}
that is based on spherical codes.
The construction will be recapped
in Section~\ref{sec:charmulerr}, and the next theorem
summarizes its properties.

\begin{theorem}[{\cite[Proposition~5]{Roth2022ITW}}]
\label{thm:rothcode-2}
There exists a linear $[n,k{=}n{-}r]$ code~$\code$ over~$\R$
with $\Gamma_2(\code) = O(n/\sqrt{r})$, whenever $r/\log n$ is bounded
away from (above)~$1$.
\end{theorem}

\subsection{Disjunct matrices}
\label{sec:disjunctmatrices}

Let $n, r \in \Z^+$ and let $D \in \Int{n}$.
An $r \times n$ matrix $H = (H_{i,j})$ over $\{ 0, 1 \}$
is called \emph{$D$-disjunct}
if the union of the supports of any $D$ columns of~$H$
does not contain the support of any other column.
In other words, for any column index $j \in \Int{n}$
and a subset $\cJ \subseteq \Int{n} \setminus \set{j}$ of
$D$ additional column indexes there is
a row index $i \in \Int{r}$ such that
$H_{i,j} = 1$ while $H_{i,j'} = 0$ for all $j' \in \cJ$.
(Equivalently, every $r\times (D+1)$ submatrix of~$H$
contains $D+1$ rows that form the identity matrix.)

A \emph{$(D,\rho)$-disjunct} matrix is a $D$-disjunct matrix whose
rows all have weights bounded from above by $\rho \in \Z^+$.

Disjunct matrices play a crucial role in the area of group testing,
which studies how to identify a set of at most $D$ positive items from
a batch of $n$ total items. The basic strategy of group testing is
to group the items into several tests, i.e., some subsets of items.
In each test, a positive outcome indicates that at least one of
the items included in this test is positive and a negative outcome
indicates that all items included are negative.
A disjunct matrix~$H$ describes a nonadaptive group testing scheme:
we use the tests to index the rows and use items to index the columns.
Then the $i$th test contains the $j$th item if and only if
$H_{i,j} = 1$. It is not very difficult to see that
the $D$-disjunct property ensures that this testing scheme can
identify all the positive items as long as their number is at most~$D$.

The first explicit construction of disjunct matrices was proposed
by Kautz and Singleton~\cite{KauSin64}. Their
construction uses a Reed--Solomon
(RS) outer code concatenated with binary
unit vectors and requires $r = O(D^2 \log_D^2 n)$ tests, which matches
the best known lower bound, $\Omega(D^2 \log_D n)$,
in~\cite{DyaRyk82},\cite{Fur96}
when $D = \Theta(n^\alpha)$ for some fixed $\alpha \in (0,1)$.
Subsequently, Porat and Rothschild~\cite{PorRot11} proposed
another explicit construction, which is similar to the Kautz--Singleton
construction but uses a code meeting the Gilbert--Varshamov (G--V) bound
as the outer code. Their construction achieves $r = O(D^2 \log n)$
and outperforms the Kautz--Singleton construction in
the regime where $D = O(\poly(\log n))$.

More recently, motivated by practical applications
in group testing and wireless communication, Inan~\emph{et al.}
investigated disjunct matrices with constraints on either
the maximal row weight (i.e, $(D,\rho)$-disjunct matrices)
or the maximal column weight~\cite{InaKaiOzg20}.
In the context of this paper, we focus on $(D,\rho)$-disjunct matrices
and demonstrate in Section~\ref{sec:codefromdismat} that
$(D,\rho)$-disjunct matrices can be used to construct
analog error-correcting codes.

Inan~\emph{et al.} first examined the Kautz--Singleton construction and
the Porat--Rothschild construction and computed
the maximum row weight~$\rho$ of the corresponding disjunct matrices.

\begin{theorem}[{\cite[Theorems~2 and~3]{InaKaiOzg20}}]
\label{thm:dismat}
The Kautz--Singleton construction yields
a $(D,\rho)$-disjunct $r \times n$ matrix with
\underline{constant} row weight $\rho = n/\sqrt{r}$ and
\[
r = O\parenv*{\parenv*{\frac{D \log n}{ \log (D \log n) }}^2}.
\]
The Porat--Rothschild construction yields
a $(D,\rho)$-disjunct $r \times n$ matrix where
$\rho = \Omega(n/D)$ and $r = O(D^2 \log n)$.
\end{theorem}

In the Porat--Rothschild construction, the number of rows,
$r = O(D^2 \log n)$, meets the lower bound $\Omega(D^2 \log_D n)$
when $D$ is fixed. The following result shows that
in a $(D,\rho)$-disjunct matrix with $r = O(\log n)$ rows
one must have $\rho = \Theta(n)$; so,
in a regime where $D$ is fixed,
both $r$ and~$\rho$ in the Porat--Rothschild construction
meet their respective lower bounds.

\begin{lemma}
\label{lem:bndsmallrow}
Let~$H$ be a $(D,\rho)$-disjunct $r \times n$ matrix,
where $r \le a \log n$ for some fixed~$a$. Then $\rho = \Theta(n)$.
\end{lemma}

\begin{proof}
Since~$H$ is $D$-disjunct, it cannot contain two identical columns
and, so, $a \ge 1$. Let $\alpha \in \parenv*{0,1/2}$ be such that
$h(\alpha) = 1/(2a)$, where $h(\cdot)$ is the binary entropy function.
Then
\[
\sum_{i=0}^{\floor{\alpha r}} \binom{r}{i} \le 2^{r h(\alpha)}
\le\sqrt{n}.
\]
Hence, there are at least $n - \sqrt{n}$ columns in~$H$ each of which
has weight at least $\alpha r$. By counting the number of~$1$s in~$H$,
we get that
\[
r \rho \ge (n-\sqrt{n})(\alpha r),
\]
which implies that $\rho \ge (n-\sqrt{n})\alpha$.
\end{proof}

Inan~\emph{et al.} proved the following generic lower bound on
the number of rows of a $(D,\rho)$-disjunct matrix.

\begin{theorem}[{\cite[Theorem~8]{InaKaiOzg20}}]
\label{thm:lowbnd}
A $(D,\rho)$-disjunct $r \times n$ matrix must satisfy
\[
r \ge
\begin{cases}
\displaystyle\frac{(D+1)n}{\rho}, & \textrm{if $\rho > D+1$,}\\
\; n ,                            & \textrm{if $\rho \le D+1$.}
\end{cases}
\]
\end{theorem}

They also modified the Kautz--Singleton construction by changing
the dimension of the outer RS code and obtained
the following result.

\begin{theorem}[{\cite[Theorem 8]{InaKaiOzg20}}]
\label{thm:condismat-1}
Let $\ell \in \Z^+$, let~$q$ be a prime power, and set
\[
n = q^{\ell+1}
\quad \textrm{and} \quad
\rho = q^\ell = n^{\ell/(\ell+1)} .
\]
Also, let $D \in \Z^+$ be such that $\ell D + 1\le q$.
The Kautz--Singleton construction yields
a $(D,\rho)$-disjunct $r \times n$ matrix
with constant row weight~$\rho$ and
\[
r = (\ell D+1) \cdot q = \frac{(\ell D + 1)n}{\rho} .
\]
\end{theorem}

Substituting $\ell = 1$ in Theorem~\ref{thm:condismat-1}
yields a construction for $\rho = \sqrt{n}$ with $r = (D+1)n/\rho$,
which, in view of Theorem~\ref{thm:lowbnd}, is optimal
with respect to~$r$. In Section~\ref{sec:conofdismat},
for any $\rho \le \sqrt{n}$ such that $n/\rho$ is a prime power,
we construct optimal $(D,\rho)$-disjunct matrices
with number of rows $r = (D+1)n/\rho$.

We end this section with a new lower bound on the number of rows of
$(D,\rho)$-disjunct matrices; this bound, in turn, will imply
that for any (fixed) $\ell \ge 2$,
the matrices in Theorem~\ref{thm:condismat-1}
are asymptotically optimal when
$D = o \bigl( n^{1/(\ell(\ell+1))} \bigr)$.
We use the following terms
(as defined in the proof of Theorem~4 in~\cite{InaKaiOzg20}).
In an $r \times n$ binary matrix $H = (\bldh_j)_{j \in \Int{n}}$,
a row $i \in [r]$ is said to be \emph{private} for
a column $j \in \Int{n}$ if row~$i$ contains a~$1$ only at column~$j$.
Similarly, a \emph{private set} for
column~$j$ is defined as a subset $\cR \subseteq \Support_0(\bldh_j)$
such that $\cR \not\subseteq \Support_0(\bldh_{j'})$
for any $j' \in \Int{n} \setminus \set{j}$.

\begin{theorem}
\label{thm:newlowbnd}
Let $n, \ell, D, \rho \in \Z^+$ be such that $\rho \ge \ell D+1$.
Any $(D,\rho)$-disjunct $r \times n$ matrix must satisfy
\[
r \ge \frac{\ell D+1}{\rho} \parenv*{n-\max\set*{\binom{r}{\ell}, \binom{2\ell}{\ell}} }.
\]
In particular, if $\max \set{\binom{r}{\ell}, \binom{2\ell}{\ell}} = o(n)$, then
\[
r \ge \frac{(\ell D+1) n}{\rho} \cdot (1-o(1)) .
\]
\end{theorem}

\begin{proof}
Let~$H$ be a $(D,\rho)$-disjunct $r \times n$ matrix
where $\rho\ge \ell D+1$. Consider the columns that have
weight${} \le D$ and denote their number by $n_1$.
Since~$H$ is $D$-disjunct, each of these columns must have
a private row; hence, $n_1 \le r$.
Remove these columns along with the corresponding private rows
and let~$H'$ be the resulting $(r-n_1)\times (n-n_1)$ matrix.
Clearly, $H'$ is $(D,\rho)$-disjunct and each column
in~$H'$ has weight${} \ge D+1 \ge \ell$.

Next, consider the columns of~$H'$ that have weight${} \le \ell D$
and denote their number by $n_2$. Since~$H'$ is $D$-disjunct,
each of these columns must have a private set of size at most~$\ell$. Note that these private sets cannot be nested. If $2\ell \leq r-n_1$, it follows from the Lubell–Yamamoto–Meshalkin inequality (see \cite{Lubell66}) that  $n_2 \le \binom{r-n_1}{\ell} \le \binom{r}{\ell}$; if $2\ell > r-n_1$, it follows from Sperner's theorem that $n_2 \le \binom{r-n_1}{\floor{(r-n_1)/2}} \le \binom{2\ell}{\ell}$. Hence, $n_2 \leq \max \set{\binom{r}{\ell}, \binom{2\ell}{\ell}}$.
We remove these $n_2$ columns from~$H'$ and count the number of~$1$s
in the resulting matrix in two ways; doing so, we get
\[
(n-n_1-n_2)(\ell D+1)\le (r-n_1)\rho,
\]
which implies that
\begin{eqnarray*}
r \rho & \ge & (n-n_2)(\ell D+1) + n_1(\rho-(\ell D+1)) \\
  & \ge & (n-n_2)(\ell D+1) \\
  &\ge & \parenv*{n-{\max\set*{\binom{r}{\ell}, \binom{2\ell}{\ell}} }}(\ell D+1).
\end{eqnarray*}
\end{proof}

Taking~$\ell$ fixed and $D = o \bigl( n^{1/(\ell(\ell+1))} \bigr)$,
we get $r^\ell = (\ell D+1)^\ell n^{\ell/(\ell+1)} = o(n)$.
Hence, for this parameter range, the construction
in Theorem~\ref{thm:condismat-1}
asymptotically attains the lower bound in Theorem~\ref{thm:newlowbnd}.

\section{The Spherical-Code Construction: Locating Multiple Errors}
\label{sec:charmulerr}

When $\hp = 2$, the spherical code construction of~\cite{Roth20}
yields a linear $[n,n-r]$ code~$\code$ over~$\R$ with redundancy
$r = \Theta(\log n)$ and with $\Gamma_2(\code) = O(n/\sqrt{r})$.
In this section, we use Proposition~\ref{prop:alcharmultcrt}
to analyze the multiple-error-correcting capability of~$\code$.
In particular, we show that for any fixed $\hp > 2$,
we still have $\Gamma_\hp(\code) = O(n/\sqrt{r})$.

We first recap the construction.
Let~$B$ be a linear $[r,\kappa,d]$ code over $\F_2$ which satisfies
the following two properties:
\begin{description}
\item[(B1)]
$B$ contains the all-one codeword, and---
\item[(B2)]
$\distance(B^\perp) > 2$.
\end{description}
Let $n = 2^{\kappa-1}$ and let $B_0$ be the set of the $n$ codewords of
$B$ whose first entry is a~$0$.
Let $H = H(B)$ be
the $r \times n$ matrix over~$\R$
whose columns are obtained from the codewords in $B_0$ by replacing
the $\zeroone$~entries by $\pm 1/\sqrt{r}$.
The code $\code(B)$ is defined as the $[n,k{\ge}n{-}r]$ code over~$\R$
with the parity-check matrix~$H$.

\begin{remark}
\label{rem:Bnontrivial}
Properties~(B1)--(B2) imply that $B$ has a generator matrix
with an all-one row and with columns that are all distinct.
This, in turn, requires that $\kappa \ge 1 + \log r$. We will in fact
assume that the latter inequality is strict (in order to have $r < n$),
in which case $d < r/2$.\qed
\end{remark}

\begin{remark}
\label{rem:BGV}
In what follows, we will also use codes~$B$ which---in addition
to satisfying properties~(B1)--(B2)---attain the G--V bound, i.e.,
\[
\frac{\kappa}{r} \ge 1 - h(d/r) ,
\]
where $h(\cdot)$ is the binary entropy function.
E.g., when $r$ is a power of~$2$, the construction of
a generator matrix of such a code~$B$
can start with the $1 + \log r$ rows
of the generator matrix of the first-order binary Reed--Muller code
(thereby guaranteeing properties~(B1)--(B2)), followed by
iterations of adding rows that are within distance${} \ge d$ from
the linear span of the already-selected rows.
(As shown in~\cite{PorRot11}, this process can be carried
out by a deterministic algorithm in time $O(2^\kappa r) = O(n r)$.)\qed
\end{remark}

The property of $\distance(B^\perp) > 2$ guarantees that any
two rows of~$H$ are orthogonal which, in turn, implies that
\[
\norm{H\bldepsilon^\transpose}_2 \le \frac{n}{\sqrt{r}} ,
\quad
\textrm{for every $\bldepsilon \in \Cube(n,1)$}.
\]
Equivalently,
\begin{equation}
\label{eq:sinnradius}
\norm{\blds}_2 \le \frac{4n^2}{r} ,
\quad
\textrm{for every $\blds \in 2 \Sinn$} ,
\end{equation}
where $\Sinn = \Sinn(H)$ and $2 \Sinn$
are as defined in~\eqref{eq:sinn}--\eqref{eq:2sinn}.
The minimum Hamming distance of~$B$ and property~(B1) jointly imply
that for any two distinct columns $\bldh_i$ and $\bldh_j$ in~$H$,
\begin{equation}
\label{eq:coherence}
\abs{\bldh_i^\transpose \cdot \bldh_j}
= \cos(\phi_{i,j}) \le 1 - \frac{2d}{r},
\end{equation}
where $\phi_{i,j}$ is the angle between $\bldh_i$ and $\bldh_j$.
Then, using geometric arguments, it is shown in~\cite{Roth20} that
\[
\Gamma_2(\code(B))
\le \frac{n/\sqrt{r}}{\min_{i \ne j} \sin(\phi_{i,j})}
\le \frac{n}{\sqrt{d(1-d/r)}} .
\]
As argued in~\cite{Roth20},
we can now select~$B$ to be a linear $[r,\kappa,d]$ code
over $\F_2$ that satisfies properties~(B1)--(B2)
with both $\kappa/r$ and $d/r$ bounded away from~$0$,
in which case the code $\code(B)$ has
$r = \Theta(\log n)$ and $\Gamma_2(\code(B)) = O(n/\sqrt{r})$.

Turning now to $\hp > 2$,
we make use of of the following concepts used in the theory of
compressed sensing~%
\cite{Bouretal},%
\cite{Candes2008},%
\cite{CanRomTao06},%
\cite{Donoho}.
Let $H = (\bldh_j)_{j \in \Int{n}}$ be an $r \times n$ matrix over~$\R$
and let $\hp \in \Int{1:n{+}1}$ and $\gamma \in \R^+$.
We say that~$H$ satisfies
\emph{the restricted isometry property (RIP)} of order~$\hp$
with constant~$\gamma$, if for every $\blde \in \Ball(n,\hp)$,
\[
(1 - \gamma) \norm{\blde}_2^2
\le \norm{H \blde^\transpose}_2^2
\le (1 + \gamma) \norm{\blde}_2^2 .
\]
In what follows we concentrate on matrices whose columns are
unit vectors, i.e., $\norm{\bldh_j}_2 = 1$ for all $j \in \Int{n}$.
For such matrices, we define the \emph{coherence} by
\[
\mu(H) \eqdef
\max_{i\ne j} \, \abs{\bldh_i^\transpose\cdot \bldh_j}.
\]

\begin{proposition}[{\cite[Proposition~1]{Bouretal}}]
\label{prop:cohtoRIP}
Let~$H$ be an $r \times n$ matrix over~$\R$
with columns that are unit vectors
and with coherence $\mu = \mu(H)$,
and let $\hp \in \Z^+$ be such that $\hp \le n$.
Then~$H$ satisfies the RIP of order~$\hp$ with constant $(\hp-1)\mu$.
\end{proposition}

Under the conditions of Proposition~\ref{prop:cohtoRIP},
for every $\blde \in \Ball(n,\hp)$ we then have
\begin{equation}
\label{eq:cohtoRIP}
\norm{H \blde^\transpose}_2^2
\ge
(1 - (\hp-1)\mu) \norm{\blde}_2^2 .
\end{equation}

\begin{theorem}
\label{thm:sphericalcodes}
Let~$B$ be a linear $[r,\kappa,d{<}r/2]$ code over $\F_2$
that satisfies properties~(B1)--(B2). Denote
\begin{equation}
\label{eq:vartheta}
\vartheta \eqdef 1 - \frac{2d}{r} ,
\end{equation}
and let $\hp \in \Z^+$ be such that $\hp \le \ceil{1/\vartheta}$.
Then
\[
\Gamma_\hp(\code(B)) \le \frac{2n}{\sqrt{r(1-(\hp-1)\vartheta)}}.
\]
In particular, if~$B$ attains the G--V bound, then
\[
\Gamma_\hp(\code(B))
\le \frac{2n}{\sqrt{r-(\hp-1) \sqrt{2 \cdot r \cdot\ln \, (2n)}}} ,
\]
for every $\hp \in \Z^+$ for which
the denominator under the outer square root is positive.
\end{theorem}

\begin{proof}
Let $H = H(B)$ be the $r \times n$ parity-check matrix
that was used to define $\code(B)$ and let $\mu = \mu(H)$.
Each column in~$H$ is a unit vector and, so, from~\eqref{eq:coherence}
we get
\begin{equation}
\label{eq:muvartheta}
\mu \le 1 - \frac{2d}{r}
\stackrel{\textrm{\scriptsize\eqref{eq:vartheta}}}{=}
\vartheta.
\end{equation}

Let
\begin{equation}
\label{eq:Delta1}
\Delta \eqdef \frac{2n}{\sqrt{r(1-(\hp-1)\vartheta)}} ,
\end{equation}
where the condition $\hp \le \ceil{1/\vartheta}$ guarantees that
$(\hp-1)\vartheta < 1$.
Also, let~$\blde$ be an arbitrary vector in
$\Ball_\Delta(n,\hp)$ (see~\eqref{eq:BsubDelta}).
For such a vector,
\begin{equation}
\label{eq:enorm}
\norm{\blde}_2 \ge \norm{\blde}_\infty > \Delta
\end{equation}
and, so,
\begin{eqnarray}
\norm*{H\blde^\transpose}_2^2
& \stackrel{\textrm{\scriptsize \eqref{eq:cohtoRIP}}}{\ge} &
(1 - (\hp-1)\mu) \norm{\blde}_2^2 \nonumber \\
\label{eq:normbnd-1}
& \stackrel{%
     \textrm{\scriptsize \eqref{eq:muvartheta}+\eqref{eq:enorm}}}{>} &
(1 - (\hp-1)\vartheta) \Delta^2
\stackrel{\textrm{\scriptsize \eqref{eq:Delta1}}}{=}
\frac{4n^2}{r}.
\end{eqnarray}
It therefore follows from~\eqref{eq:sinnradius} that
\[
H\blde^\transpose \notin 2 \Sinn ,
\]
and by Proposition~\ref{prop:alcharmultcrt}
we thus conclude that $\Gamma_\hp(\code(B)) \le \Delta$.

If~$B$ attains the G--V bound, then
\begin{equation}
\label{eq:entropfunc}
\frac{\kappa}{r}
\ge 1 - h(d/r) = 1 - h(1/2-(\vartheta/2)) > \vartheta^2/c,
\end{equation}
where $c = 2\ln 2$. From $n = 2^{\kappa-1}$ we then get
\[
\log \, (2n) = \kappa > r\cdot \vartheta^2/c,
\]
or
\[
\vartheta < \sqrt{\frac{c \cdot \log \, (2n)}{r}}
= \sqrt{\frac{2 \cdot \ln \, (2n)}{r}} \; .
\]
Hence, in this case,
\begin{eqnarray*}
\Gamma_\hp(\code(B)) \le \Delta
& = & \frac{2n}{\sqrt{r(1-(\hp-1)\vartheta)}} \\
& < & \frac{2n}{\sqrt{r-(\hp-1) \sqrt{2 \cdot r \cdot\ln \, (2n)}}} \; .
\end{eqnarray*}
\end{proof}

The next lemma (which is proved in the Appendix)
presents an alternative to the bound~\eqref{eq:cohtoRIP}
that leads to some improvement on Theorem~\ref{thm:sphericalcodes}.
For $\vartheta \in (0,1)$
and a positive integer $\hp \le \ceil{1/\vartheta}$,
we introduce the notation
\[
\eta_\hp(\vartheta) \triangleq
\frac{1}{1/\vartheta + 2 - \hp} .
\]

\begin{remark}
\label{rem:RIP-alt}
In the range $1 \le \hp \le \ceil{1/\vartheta}$
we have $\eta_\hp(\vartheta) < 1$.
Also, it is easy to verify by differentiation that
in that range of~$\vartheta$ (when $\hp$ is assumed to be fixed),
the mapping
$\vartheta \mapsto (1 + \vartheta) \parenv*{1 - \eta_\hp(\vartheta)}$
is non-increasing.\qed
\end{remark}

\begin{lemma}
\label{lem:RIP-alt}
Let~$H$ be an $r \times n$ matrix over~$\R$
with columns that are unit vectors and with coherence $\mu = \mu(H)$,
and let $\hp \in \Z^+$ be such that
$\hp \le \min \set*{\ceil{1/\mu},n}$.
Then for every $\blde \in \Ball(n,\hp)$,
\[
\norm{H \blde^\transpose}_2^2
\ge
(1 + \mu) \parenv*{1 - \eta_\hp(\mu)} \norm{\blde}_\infty^2 .
\]
\end{lemma}

\begin{theorem}
\label{thm:sphericalcodes-improved}
Under the conditions of Theorem~\ref{thm:sphericalcodes},
\[
\Gamma_\hp(\code(B))
\le
\frac{2n}{\sqrt{r \cdot (1 + \vartheta)(1 - \eta_\hp(\vartheta))}} \; .
\]
\end{theorem}

\begin{proof}
Let
\begin{equation}
\label{eq:Delta2}
\Delta \eqdef
\frac{2n}{\sqrt{r \cdot (1 + \vartheta)(1 - \eta_\hp(\vartheta))}} \; .
\end{equation}
Referring to the proof of
Theorem~\ref{thm:sphericalcodes}, by applying
Lemma~\ref{lem:RIP-alt}
we can replace~\eqref{eq:normbnd-1} by
\begin{eqnarray*}
\norm*{H \blde^\transpose}_2^2
& \ge &
(1 + \mu) \parenv*{1 - \eta_\hp(\mu)}
\norm{\blde}_\infty^2 \\
& \!\!\!\!\stackrel{\textrm{\scriptsize
  \eqref{eq:muvartheta}+\eqref{eq:enorm}+Remark~\ref{rem:RIP-alt}}}{>}
  \!\!\!\! &
(1 + \vartheta) \parenv*{1 - \eta_\hp(\vartheta)} \Delta^2
\stackrel{\textrm{\scriptsize \eqref{eq:Delta2}}}{=}
\frac{4n^2}{r}.
\end{eqnarray*}
And as in that proof, we then conclude that
$\Gamma_\hp(\code(B)) \le \Delta$.
\end{proof}

When $1 < \hp < 1+1/\vartheta$, we have
\[
1 - \vartheta(\hp-1) < (1+\vartheta)(1-\eta_\hp(\vartheta))
\]
and so, Theorem~\ref{thm:sphericalcodes-improved} is stronger
than Theorem~\ref{thm:sphericalcodes}.
The improvement of Theorem~\ref{thm:sphericalcodes-improved}
is seen best when~$\hp$ is close to $\ceil{1/\vartheta}$.\footnote{%
This means that given~$n$ and~$r$,
we select~$\hp$ to be close to the largest possible and analyze which
values of $\Gamma_\hp(\code(B))$ can then be attained.}
For example, when $\hp = 1/\vartheta$,
Theorem~\ref{thm:sphericalcodes} yields the upper bound
\[
\Gamma_\hp(\code(B)) \le \sqrt{\hp} \cdot \frac{2n}{\sqrt{r}} ,
\]
while from Theorem~\ref{thm:sphericalcodes-improved} we get:
\begin{equation}
\label{eq:largem}
\Gamma_\hp(\code(B))
\le \sqrt{\frac{2\hp}{\hp+1}} \cdot \frac{2n}{\sqrt{r}}
< \sqrt{8} \cdot \frac{n}{\sqrt{r}} .
\end{equation}
In fact, \eqref{eq:largem} is the bound we get
in Theorem~\ref{thm:sphericalcodes}
when we reduce~$\hp$ (by almost half) to $1/(2\vartheta) + 1$
while, for this~$\hp$, Theorem~\ref{thm:sphericalcodes-improved} yields
\[
\Gamma_\hp(\code(B))
\le \sqrt{\frac{2\hp}{2\hp-1}} \cdot \frac{2n}{\sqrt{r}} .
\]
When $\hp \ll 1/\vartheta$,
the upper bounds in both theorems approach $2n/\sqrt{r}$.

\begin{corollary}
\label{cor:newasymbound}
For any $n, \hp \in \Z^+$
there exists a linear $[n,k{\ge}n{-}r]$ code~$\code$ over~$\R$ with
\[
r = 2 \hp^2 \ceil{\ln \, (2n)}
\]
and
\[
\Gamma_\hp(\code) < \sqrt{8} \cdot \frac{n}{\sqrt{r}}
\le
\frac{2 n}{\hp \sqrt{\ln \, (2n)}} .
\]
\end{corollary}

\begin{proof}
Write $\vartheta = 1/\hp$
and let~$B$ be a linear $[r,\kappa,d]$ code over $\F_2$
that satisfies properties~(B1)--(B2) with parameters
\[
r = 2 \hp^2 \ceil{\ln \, (2n)}
\quad \textrm{and} \quad
d = \hp (\hp - 1) \ceil{\ln \, (2n)} ,
\]
in which case
\[
\vartheta \eqdef 1 - \frac{2d}{r} = \frac{1}{\hp} .
\]
Indeed, by the G--V bound~\eqref{eq:entropfunc}, such a code exists with
dimension
\[
\kappa > \frac{r \cdot \vartheta^2}{2 \ln 2} \ge \log \, (2n)
\]
and, so, the respective code $\code(B)$ has length $2^{\kappa-1} > n$
and can be shortened to form a linear $[n,k{\ge}n{-}r]$ code~$\code$
over~$\R$.
Finally, since $\hp = 1/\vartheta$,
we get from~\eqref{eq:largem} that
$\Gamma_\hp(\code)\le \Gamma_\hp(\code(B)) < \sqrt{8} \cdot n/\sqrt{r}$.
\end{proof}

\begin{remark}
\label{rem:newasymbound-1}
The last corollary
is non-vacuous when $\hp = O \bigl( \sqrt{n / \log n} \bigr)$
(otherwise we have $r > n$).
When $\hp = 2$, the corollary coincides
with Theorem~\ref{thm:rothcode-2}.\qed
\end{remark}

\begin{remark}
\label{rem:newasymbound-2}
In Corollary~\ref{cor:newasymbound},
we can make~$r$ grow more slowly with~$\hp$
at the expense of a faster growth with $\log n$,
while keeping the same upper bound
$\Gamma_\hp(\code) \le \sqrt{8} \cdot n/\sqrt{r}$.
Specifically, in the proof, we take~$B$ to be the dual of
an extended binary BCH primitive code~\cite[p.~280]{MacSlo78},
or as a concatenation of a RS outer code with
the first-order binary Reed--Muller code.
In both cases we have, for a parameter $t \in \Z^+$,
\[
\vartheta = 1 - \frac{2d}{r} = O \parenv*{\frac{t}{\sqrt{r}}}
\quad \textrm{and} \quad
\kappa = \Theta(t \log r) ,
\]
i.e.,
\[
\vartheta = O \parenv*{\frac{\kappa}{\sqrt{r} \cdot \log r}} .
\]
Substituting
$\kappa = \ceil{\log \, (2n)}$ and $\vartheta = 1/\hp$ then yields
\[
r \log^2 r = O \parenv*{\hp^2 \log^2 n} ,
\]
which is non-vacuous when $\hp = O(\sqrt{n})$.\qed
\end{remark}

\section{Code Construction Based on Disjunct Matrices}
\label{sec:codefromdismat}

In this section, we study the relationship between
analog error-correcting codes and disjunct matrices.
Specifically, we consider linear codes over~$\R$ with
parity-check matrices that are $(D,\rho)$-disjunct:
we first study their properties
(Theorem~\ref{thm:disjtoerrloc})
and then propose decoding algorithms for these codes.

\begin{theorem}
\label{thm:disjtoerrloc}
Let~$H$ be
\an~$(\hp{-}1,\rho)$-disjunct $r \times n$ matrix,
for some $\hp, \rho \in \Int{1:n{+}1}$,
and let~$\code$ be the linear $[n,k{\ge}n{-}r]$ code over~$\R$
that has~$H$ as a parity-check matrix. Then
\[
\Gamma_\hp(\code) \le 2\rho.
\]
\end{theorem}

\begin{proof}
We show that $\Delta = 2\rho$ is contained
in the minimand set in~\eqref{eq:alcharmultcrt};
the result will then follow from Proposition~\ref{prop:alcharmultcrt}.
Given any vector $\blde=(e_j)_{j \in \Int{n}} \in \Ball_\Delta(n,\hp)$,
write $\cJ = \Support_0(\blde)$
and let $t \in \cJ$ be a position at which $\abs{e_t} > \Delta$.
Since~$H$ is $(\hp{-}1)$-disjunct and $\abs{\cJ} \le \hp$,
there is a row index $i \in \Int{r}$ such that
$(H_i)_\cJ$ contains a~$1$ only at position~$t$.
Therefore,
\begin{equation}
\label{eq:disjtoerrloc1}
\abs*{H_i \blde^\transpose} = \abs{e_t} > \Delta = 2\rho .
\end{equation}
On the other hand, since $\weight(H_i) \le \rho$,
for every $\bldepsilon \in \Cube(n,2)$ we have
$\abs{H_i \bldepsilon^\transpose} \le 2\rho$,
namely,
\begin{equation}
\label{eq:disjtoerrloc2}
\abs{s_i} \le 2\rho ,
\quad \textrm{for every $\blds = (s_v)_{v \in \Int{r}} \in 2 \Sinn$} .
\end{equation}
By~\eqref{eq:disjtoerrloc1} and~\eqref{eq:disjtoerrloc2}
we get that $H \blde^\transpose \notin 2 \Sinn$,
thus establishing that $\Delta = 2\rho$ is contained
in the minimand in~\eqref{eq:alcharmultcrt}.
\end{proof}

Combining Theorem~\ref{thm:disjtoerrloc} with
Theorem~\ref{thm:condismat-1}, we obtain the following result.

\begin{corollary}
\label{cor:code-1}
Let $\ell \in \Z^+$, let~$q$ be a prime power,
and set $n = q^{\ell+1}$.
Then for any positive integer $\hp \le \ceil{q/\ell}$
there is an explicit construction of
a linear $[n,k{\ge}n{-}r]$ code~$\code$ over~$\R$ such that
\[
r = (\ell \hp - \ell + 1) q
\]
and
\[
\Gamma_\hp(\code)\le 2 q^\ell
= \frac{2(\ell \hp - \ell + 1)n}{r} .
\]
In particular, by taking $\ell = 1$,
for any $\hp \le \sqrt{n}$
one can obtain a linear code~$\code$ with
\[
r = \hp \sqrt{n} \quad \textrm{and} \quad
\Gamma_\hp(\code) \le 2 \sqrt{n} = \frac{2 \hp n}{r} .
\]
\end{corollary}

It is worth noting that when $\hp = 2$, the bound
$\Gamma_2(\code) \le 4n/r$ coincides with the one in
Theorem~\ref{thm:rothcode} (although~$r$ in that theorem can take
multiple values, including values that are smaller than $2 \sqrt{n}$).
We also note that in Corollary~\ref{cor:code-1},
we have
$r = \Theta(\hp n^\alpha)$ and $\Gamma_\hp(\code) = O(\hp n/r)$,
for certain (fixed) $\alpha \in (0,1/2]$
and infinitely many values of~$n$.
In Section~\ref{sec:conofdismat},
we present a construction of disjunct matrices
which produce codes with
similar dependence of~$r$ and~$\Gamma_\hp(\cdot)$
on~$n$ and~$\hp$, yet for $\alpha \in [1/2,1)$.

Next, we compare the construction of
Corollary~\ref{cor:newasymbound}
with the case $\ell = 1$ in Corollary~\ref{cor:code-1}
(as this case yields the slowest growth of~$r$ with~$n$).
For the former we have
$\Gamma_\hp(\code) \le \sqrt{8} \cdot n/\sqrt{r}$,
while for the latter $\Gamma_\hp(\code) = \sqrt{n}$, which is smaller
since $r < n$. Yet the construction of Corollary~\ref{cor:code-1}
requires $r = \hp \sqrt{n}$, which can match the redundancy,
$2 \hp^2 \ceil{\ln \, (2n)}$,
in Corollary~\ref{cor:newasymbound} only when
\[
\hp = \Omega \parenv*{{\sqrt{n}}/{\log n}}
\]
(still, by Remark~\ref{rem:newasymbound-1},
this range partially overlaps with the range of~$\hp$ for which
the codes in Corollary~\ref{cor:newasymbound} are realizable).

\begin{remark}
\label{rem:disjtoerrloc}
The construction of
Theorem~\ref{thm:disjtoerrloc},
when applied with the Porat--Rothschild disjunct matrices in
Theorem~\ref{thm:dismat},
yields $r = O(\hp^2 \log n)$
(i.e., a similar guarantee to that in Corollary~\ref{cor:newasymbound})
yet with $\Gamma_\hp(\code) = \Omega(n/\hp)$,
which is $\Omega(\sqrt{\log n})$ times larger than
the respective value in Corollary~\ref{cor:newasymbound}.\qed
\end{remark}


In the remainder of this section, we present decoders
for linear codes with parity-check matrices that are
$(\hp{-}1,\rho)$-disjunct.
In Subsection~\ref{sec:decoder-generic} we present a decoder
for the generic case, yet its complexity is $O(r \hp n^\hp)$,
i.e., polynomial only when~$\hp$ is fixed.
A much more efficient algorithm is presented
in Subsection~\ref{sec:decoder-both-constrained},
yet under the additional assumption that the column weights in
the parity-check matrix are also constrained.

\subsection{Decoder for the generic disjunct construction}
\label{sec:decoder-generic}

Our first decoder, denoted by~$\Decoderb$, is presented in
Algorithm~\ref{alg:decoder-2}.

\begin{algorithm}[ht]
\begin{algorithmic}
\caption{Decoder~$\Decoderb$ for codes from disjunct matrices}
\label{alg:decoder-2}
\State{$\triangleright$
  $H = (H_{i,j})$ is
  \an~$(\hp{-}1,\rho)$-disjunct $r \times n$ matrix}
\State{$\triangleright$
  $\tau, \sigma \in \Z_{\ge 0}$ are such that $2\tau + \sigma = \hp$}
\State{\textbf{Input}: vector $\bldy \in \R^n$}
\State{\textbf{Output}: subset $\Decoderb(\bldy) \subseteq \Int{n}$}
\State{}
\State{Set $\Lambda = \set{(\cT,\cJ) \,:\,
   \varnothing \ne \cT \subseteq \cJ \subseteq \Int{n}
   \; \textrm{and} \; \abs{\cJ} \le \tau+\sigma}$}
\State{For each $(\cT,\cJ) \in \Lambda$, let
\[
\cR(\cT,\cJ) = \set*{i \in [r] \,:\,
   \weight((H_i)_\cT) = \weight((H_i)_\cJ) = 1}
\]}
\State{$\Decoderb(\bldy)\gets \varnothing$}
\State{$\blds = (s_i)_{i \in \Int{r}} \gets H \bldy^\transpose$}
\While{$\exists (\cT,\cJ)\in \Lambda$ s.t.\ $\abs{s_i} > \rho$
  for all $i \in \cR(\cT,\cJ)$}
  \State{$\Decoderb(\bldy)\gets \Decoderb(\bldy) \cup \cT$}
  \State{$\Lambda \gets \Lambda \setminus \set{(\cT,\cJ)}$}
\EndWhile
\State{\Return{$\Decoderb(\bldy)$}}
\end{algorithmic}
\end{algorithm}

\begin{theorem}
\label{thm:disjdecoder}
Let~$\code$ be a code as in Theorem~\ref{thm:disjtoerrloc}.
Then the mapping $\Decoderb : \R^n \rightarrow 2^{\Int{n}}$
that is defined by Algorithm~\ref{alg:decoder-2} is
a $(\tau,\sigma)$-decoder for $(\code, 2\rho:1)$.
\end{theorem}

\begin{proof}
Assume a received (read) vector
\[
\bldy = \bldc + \blde + \bldepsilon ,
\]
where $\bldc \in \code$,
$\bldepsilon \in \Cube(n,1)$,
and $\blde \in \Ball(n,\tau + \sigma)$.

We first show that
\begin{equation}
\label{eq:decoder-1}
\Support_{2\rho}(\blde) \subseteq \Decoderb(\bldy).
\end{equation}
Take $\cT = \Support_{2\rho}(\blde)$
and $\cJ = \Support_0(\blde)$.
Then for every $i \in \cR(\cT,\cJ)$, since $\weight((H_i)_\cT) = 1$
and $(H_i)_{\cJ \setminus \cT} = \Zero$, we have
\begin{equation}
\label{eq:si}
\abs{s_i}
= \abs{H_i \blde^\transpose + H_i \bldepsilon^\transpose}
\ge
{\underbrace{\abs{H_i \blde^\transpose}}_{{} > 2\rho}}
-
\underbrace{\abs{H_i \bldepsilon^\transpose}}_{{} \le \rho}
> 2\rho - \rho = \rho.
\end{equation}
Hence, $(\cT,\cJ)$ passes the check in the while loop and, so,
the set $\cT$ is joined into $\Decoderb(\bldy)$,
thereby establishing~\eqref{eq:decoder-1}.

Next, we assume that $\weight(\blde) \le \tau$ and show that
\begin{equation}
\label{eq:decoder-2}
\Decoderb(\bldy)\subseteq \Support_0(\blde).
\end{equation}
Write $\cK = \Support_0(\blde)$; then $\abs{\cK}\le \tau$.
Let $(\cT,\cJ)$ be a pair in~$\Lambda$ that passes the check
in the while loop, i.e.,
$\abs{s_i} > \rho$ for all $i \in \cR(\cT,\cJ)$.
We claim that $\cT \subseteq \cK$.
Otherwise, take a $t \in \cT \setminus \cK$.
Since~$H$ is $(\hp{-}1)$-disjunct and
\[
\abs{\cJ \cup \cK} \le
\abs{\cJ} + \abs{\cK} \le (\tau + \sigma) + \tau = \hp,
\]
there is a row index $i \in [r]$ such that
$H_{i,t} = 1$ and $H_{i,j} = 0$
for all $j \in \parenv*{\cJ \cup \cK} \setminus \set{t}$.
Then $\weight((H_i)_\cT) = \weight((H_i)_\cJ) = 1$
and, so, $i \in \cR(\cT,\cJ)$.
On the other hand, we also have $(H_i)_\cK = \Zero$, from which we get
\[
\abs{s_i}
= \abs{
{\underbrace{H_i \blde^\transpose}_0}
+ H_i \bldepsilon^\transpose}
\le \rho .
\]
Yet this means that the pair $(\cT,\cJ)$ does \emph{not} pass
the check in the while loop, thereby reaching a contradiction.
We conclude that when $\weight(\blde) \le \tau$,
any set~$\cT$ that is joined into $\Decoderb(\bldy)$
in the while loop is a subset of $\cK = \Support_0(\blde)$,
thus establishing~\eqref{eq:decoder-2}.

Eqs.~\eqref{eq:decoder-1} and \eqref{eq:decoder-2}, in turn, imply that
the function~$\Decoderb$ in Algorithm~\ref{alg:decoder-2}
satisfies conditions~(D2) and~(D1), respectively,
in the definition of a $(\tau,\sigma)$-decoder for $(\code, 2\rho:1)$.
\end{proof}

We note that
\[
\abs{\Lambda}
= \sum_{j=1}^{\tau+\sigma} \binom{n}{j} (2^j - 1) = O(n^{\tau+\sigma})
= O(n^\hp) .
\]
Given a pair $(\cT,\cJ) \in \Lambda$, checking the conditions in
the while loop of Algorithm~\ref{alg:decoder-2} can be done in
$O(r \hp)$ time.

\subsection{Decoder when columns in $H$ are also weight-constrained}
\label{sec:decoder-both-constrained}

Let~$\code$ be a code as in Theorem~\ref{thm:disjtoerrloc}
and~$w$ be a positive integer.
We next present a more efficient
$(\tau,\sigma)$-decoder for $(\code,2\rho:1)$
under the following two additional conditions on~$H$:
\begin{description}
\item[(H1)]
Every row of~$H$ has weight at least~$2$.
\item[(H2)]
Every column of~$H$ has weight at most~$w$.
\end{description}

Condition~(H1) is not really limiting:
the case where~$H$ contains rows of weight~$1$ is degenerate,
as then there are positions on which all
the codewords in~$\code$ are identically~$0$
(and, thus, these coordinates can be ignored, thereby reducing
the decoding to a shorter code).
In Section~\ref{sec:conofdismat}, we present constructions
of $(D,\rho)$-disjunct matrices that satisfy conditions~(H1)--(H2).

We will use the following lemma.

\begin{lemma}
\label{lem:m-1disjunct}
Let~$H$ be
\an~$(\hp{-}1)$-disjunct $r\times n$ matrix
that satisfies condition~(H1).
Given any nonempty subset $\cJ \subseteq \Int{n}$
of size $\abs{\cJ} \le \hp$,
for every column index $j \in \cJ$
there exist at least $\hp+1-\abs{\cJ}$ nonzero rows in
the submatrix $(H)_\cJ$
that contain a~$1$ only at column~$j$.
\end{lemma}

\begin{proof}
The proof is by backward induction on~$\abs{\cJ}$,
with the induction base, $\abs{\cJ} = \hp$, following from
the definition of a $(\hp{-}1)$-disjunct matrix.

Turning to the induction step, suppose that
$0 < \abs{\cJ}\le \hp-1$
and let~$j$ be any column index in~$\cJ$.
By the disjunct property, there exists a row index
$i \in \Int{r}$ such that
$(H_i)_\cJ$ contains a~$1$ only at position~$j$.
By condition~(H1), there is at least one
index $j' \in \Int{n} \setminus \cJ$ for which $H_{i,j'} = 1$.
Letting $\cJ' = \cJ \cup \set{j'}$,
by the induction hypothesis
there are at least
$\hp + 1 - \abs{\cJ'} = \hp - \abs{\cJ}$
nonzero rows in $(H)_{\cJ'}$
that contain a~$1$ only at column~$j$; clearly, none of these
rows is indexed by~$i$ since $(H_i)_{\cJ'}$ contains
two~$1$s. Altogether there are at least
$\hp+1-\abs{\cJ}$ nonzero rows in $(H)_\cJ$
that contain a~$1$ only at column~$j$.
\end{proof}

\begin{remark}
\label{rem:w>=m}
Applying Lemma~\ref{lem:m-1disjunct} with $\abs{\cJ} = 1$
implies that the weight of every column in~$H$ must be
at least~$\hp$
(recall that we have used this fact in
the proof of Theorem~\ref{thm:newlowbnd}).
Hence, $(\lambda{-}1)$-disjunct matrices
can satisfy conditions~(H1) and~(H2)
only when $w \ge \hp$.\qed
\end{remark}

Given $\rho \in \R^+$
and a vector $\blds = (s_i)_{i \in \Int{r}} \in \R^r$
(such as a syndrome that is computed with respect to~$H$),
we let $\bldchi_\rho(\blds)$ be the real row vector
$(\chi_i)_{i \in \Int{r}} \in \set{0, 1}^r$
whose entries are given by
\[
\chi_i
= \begin{cases}
0, & \textrm{if $\abs{s_i} \le \rho$}, \\
1, & \textrm{otherwise}. \\
\end{cases}
\]

\begin{theorem}
\label{thm:disjunctdecoder}
Let~$\code$ be a code as in Theorem~\ref{thm:disjtoerrloc}
and suppose that $H$ also satisfies conditions~(H1)--(H2).
For nonnegative integers~$\tau$ and~$\sigma$ such that
\begin{equation}
\label{eq:tausigma}
2\tau + \sigma \le 2\hp - w \; (\le \hp) ,
\end{equation}
let $\Decodert : \R^n \rightarrow 2^{\Int{n}}$
be defined for every $\bldy \in \R^n$ by
\begin{equation}
\label{eq:disjunctdecoder}
\Decodert(\bldy) \eqdef
\Support_{\hp-\tau-\sigma} \parenv*{\bldchi_\rho (\blds) H} ,
\end{equation}
where $\blds = H \bldy^\transpose$.
Then~$\Decodert$ is a $(\tau,\sigma)$-decoder for $(\code,2\rho:1)$.
\end{theorem}

\begin{proof}
Assume a received (read) vector
\[
\bldy = \bldc + \blde + \bldepsilon ,
\]
where $\bldc \in \code$,
$\bldepsilon \in \Cube(n,1)$,
and $\blde \in \Ball(n,\tau + \sigma)$.

We first show that
\begin{equation}
\label{eq:case-1}
\Support_{2\rho}(\blde) \subseteq \Decodert(\bldy) .
\end{equation}
Take $\cJ = \Support_0(\blde)$
and let $j \in \Support_{2\rho}(\blde) \; (\subseteq \cJ)$.
By Lemma~\ref{lem:m-1disjunct} we get
that the submatrix $(H)_\cJ$ contains at least
\begin{equation}
\label{eq:overlap1}
\hp + 1 - \abs{\cJ} \ge \hp + 1 - \tau - \sigma
\end{equation}
rows with a~$1$ only at column~$j$.
Denoting by~$\cR$ the set of indexes of these rows,
for every $i \in \cR$,
the respective entry $s_i$ in the syndrome~$\blds$ satisfies:
\[
\abs{s_i} = \abs{H_i \blde^\transpose + H_i \bldepsilon^\transpose}
\ge \abs{H_i \blde^\transpose} - \abs{H_i \bldepsilon^\transpose}
> \rho
\]
(similarly to~\eqref{eq:si}).
It follows that the respective entry, $\chi_i$, in
$\bldchi_\rho(\blds)$ equals~$1$ and, so,
the supports of $\bldchi_\rho(\blds)$
and the column~$\bldh_j$ in~$H$
overlap on at least $\abs{\cR}$ positions. Hence,
\[
\bldchi_\rho(\blds) \cdot \bldh_j
\ge \abs{\cR}
\stackrel{\textrm{\scriptsize \eqref{eq:overlap1}}}{\ge}
\hp + 1 - \tau - \sigma ,
\]
i.e.,
$j \in \Support_{\hp-\tau-\sigma} \parenv*{\bldchi_\rho (\blds) H}
\eqdef \Decodert(\bldy)$.
We conclude that
\[
j \in \Support_{2\rho}(\blde)
\; \Longrightarrow \;
j \in \Decodert(\bldy) ,
\]
thereby establishing~\eqref{eq:case-1}.

Next, we assume that $\weight(\blde) \le \tau$ and show that
\begin{equation}
\label{eq:case-2}
\Decodert(\bldy) \subseteq \Support_0(\blde) .
\end{equation}
Write $\cK = \Support_0(\blde)$
and let $j \in \Int{n} \setminus \cK$.
Lemma~\ref{lem:m-1disjunct}, now applied with
$\cJ = \cK \cup \set{j}$,
implies that the submatrix $(H)_\cJ$
contains at least
\begin{equation}
\label{eq:overlap2}
\hp + 1 - \abs{\cJ} \ge \hp - \tau
\end{equation}
rows with a~$1$ only at column~$j$.
Letting~$\cR$ be the set of indexes of these rows, for every $i \in \cR$
we then have $(H_i)_\cK = \Zero$
and, so, the respective entry in the syndrome~$\blds$ satisfies:
\[
\abs{s_i}
= \abs{
{\underbrace{H_i \blde^\transpose}_0}
+ H_i \bldepsilon^\transpose}
\le \rho ,
\]
namely, $\chi_i = 0$.
Hence, the number of positions on which
the supports of $\bldchi_\rho(\blds)$ and~$\bldh_j$
overlap is at most
\[
\weight(\bldh_j) - \abs{\cR}
\stackrel{\textrm{\scriptsize (H2)+\eqref{eq:overlap2}}}{\le}
w - (\hp - \tau)
\stackrel{\textrm{\scriptsize \eqref{eq:tausigma}}}{\le}
\hp - \tau - \sigma
\]
and, so,
\[
(0 \le) \;
\bldchi_\rho(\blds) \cdot \bldh_j \le \hp - \tau - \sigma ,
\]
i.e.,
$j \notin \Support_{\hp-\tau-\sigma} \parenv*{\bldchi_\rho (\blds) H}
\eqdef \Decodert(\bldy)$.
We conclude that when $\weight(\blde) \le \tau$,
\[
j \notin \Support_0(\blde)
\; \Longrightarrow \;
j \notin \Decodert(\bldy) ,
\]
thereby establishing~\eqref{eq:case-2}.

Eqs.~\eqref{eq:case-1} and~\eqref{eq:case-2}, in turn, imply
that the function $\bldy \mapsto \Decodert(\bldy)$
defined in~\eqref{eq:disjunctdecoder}
is a $(\tau,\sigma)$-decoder for $(\code,2\rho:1)$.
\end{proof}

The decoder~\eqref{eq:disjunctdecoder} is easy to compute:
it consists of a multiplication of~$H$ to the right by~$\bldy$
to obtain the syndrome~$\blds$, and then to the left
by a binary vector which is a quantized copy of~$\blds$.
Since~$H$ is a $\zeroone$ matrix whose rows and columns
have limited weights (at most~$\rho$ and~$w$, respectively),
the decoding requires less than $2 \min \set{r \rho, w n}$
real additions.

We note that the condition~\eqref{eq:tausigma}
(which was used in our analysis), is generally
stricter than the condition $2\tau + \sigma \le \hp$ which,
by Theorems~\ref{thm:disjtoerrloc} and~\ref{thm:disjdecoder},
is sufficient
for having a $(\tau,\sigma)$-decoder for $(\code,2\rho:1)$.
These two conditions coincide when $w = \hp$, and this case
is characterized in the next lemma.

\begin{lemma}
\label{lem:w=m}
Let $H = (\bldh_j)_{j \in \Int{n}}$ be
\an~$(\hp{-}1)$-disjunct $r\times n$ matrix that satisfies
conditions~(H1)--(H2) with $w = \hp$.
Then the following holds.
\begin{description}
\item[M1)]
Every column of~$H$ has weight (exactly)~$\hp$.
\item[M2)]
The supports of every two distinct columns of~$H$
intersect on at most one coordinate.
\end{description}
Equivalently, for every $j \ne j'$ in $\Int{n}$:
\[
\norm{\bldh_j}_2 = \sqrt{\hp}
\quad \textrm{and} \quad
\abs{\bldh_j^\transpose \cdot \bldh_{j'}} \le 1
\]
(and, thus, the columns of~$H$ constitute a spherical code).
\end{lemma}

\begin{proof}
Condition~(M1) follows from Lemma~\ref{lem:m-1disjunct}
when applied with $\abs{\cJ} = 1$ (see Remark~\ref{rem:w>=m}),
and condition~(M2) follows from applying the lemma with $\abs{\cJ} = 2$.
\end{proof}

We end this section by presenting a simple decoder for
the detection-only case, i.e., $\tau = 0$.
In this case, we actually do not need conditions~(H1)--(H2),
and we can handle any $\sigma \le \hp$.

\begin{theorem}
\label{thm:detection-only}
Let~$\code$ be a code as in Theorem~\ref{thm:disjtoerrloc}
and let $\Decoderh : \R^n \rightarrow \set{\varnothing, \failure}$
be defined by
\[
\Decoderh(\bldy)
= \begin{cases}
\varnothing, & \textrm{if $\bldchi_\rho(\blds) = \Zero$}, \\
\failure,    & \textrm{otherwise}, \\
\end{cases}
\]
where $\blds = H \bldy^\transpose$.
Then~$\Decoderh$ is a $(0,\hp)$-decoder for $(\code,2\rho:1)$.
\end{theorem}

\begin{proof}
Condition~(D1) pertains only
to the error vector $\blde = \Zero$, in which case
\[
\norm{\blds}_\infty
= \norm{
{\underbrace{H \blde^\transpose}_{\Zero}}
+ H \bldepsilon^\transpose}_\infty
\le \rho .
\]
Consequently, $\bldchi_\rho(\blds) = \Zero$ and we
have $\Decoderh(\bldy) = \varnothing$, as required.

As for condition~(D2), we have $\Decoderh(\bldy) \ne \failure$
only when $\bldchi_\rho(\blds) = \Zero$.
Now, in the proof of Theorem~\ref{thm:disjunctdecoder},
we have established~\eqref{eq:case-1}
without using conditions~(H1)--(H2);
hence, we can apply~\eqref{eq:case-1}
to conclude that
\[
\Support_{2\rho}(\blde)
\subseteq \Support_{\hp-\tau-\sigma}
\bigl(\bldchi_\rho (\blds) H \bigr) \bigm|_{\tau=0,\sigma=\hp}
= \varnothing = \Decoderh(\bldy) ,
\]
as required.
\end{proof}

\section{Constructions of Disjunct Matrices with Weight-Constrained
Rows and Columns}
\label{sec:conofdismat}

In this section, we present several constructions for
$(D,\rho)$ disjunct matrices which satisfy conditions~(H1)--(H2) with
$w = D+1$. Our constructions are based on combinatorial designs.
We start by recalling several definitions.

Let $t, r, s \in \Z^+$ be such that $r \ge s \ge t$.
A \emph{$t$-$(r,s,1)$ packing design} is a pair $(X,\Block)$,
where $X$ is a set of $r$~elements (called \emph{points})
and~$\Block$ is a collection of $s$-subsets
(called \emph{blocks}) of~$X$,
such that every $t$-subset of~$X$ is contained in at most one block.
Furthermore, a packing design is called \emph{resolvable}
if its blocks can be partitioned into sets (\emph{parallel classes})
$\cP_0,\cP_1,\ldots,\cP_{\rho-1}$ such that each point is contained
in exactly one block in each $\cP_i$.

The \emph{incidence matrix} of packing design $(X,\Block)$ is
an $|X| \times |\Block|$
binary matrix $H = (H_{x,\beta})$ whose rows and columns are indexed
by the elements of~$X$ and~$\Block$, respectively, 
and for each $x \in X$ and $\beta \in \Block$,
\[
H_{x,\beta} =
\begin{cases}
    1, & \textrm{if $x \in \beta$,}\\
    0, & \textrm{if $x \notin \beta$}.
\end{cases}
\]

\begin{proposition}
\label{prop:incmatpacking}
Let $(X,\Block)$ be a resolvable $t$-$(r,s,1)$ packing design with
$\rho$~parallel classes. Then its incidence matrix~$H$ is
a $D$-disjunct matrix with constant row weight~$\rho$,
where $D = \floor{(s-1)/(t-1)}$.
\end{proposition}

\begin{proof}
Write $H = (\bldh_\beta)_{\beta \in \Block}$
and let $\beta_0, \beta_1, \ldots, \beta_D$ be
arbitrary $D+1$ blocks in~$\Block$.
Since $(X,\Block)$ is a $t$-$(r,s,1)$ packing design,
every two blocks of $\Block$ have at most $t-1$ common points.
Then $\abs*{\beta_0 \cap \beta_j} \le t-1$
for all $1 \le j \le D$ and, so,
\[
\abs*{\beta_0 \cap \parenv*{\cup_{j=1}^D \beta_j}}
\le \sum_{j=1}^D \abs*{\beta_0\cap \beta_j} \le D(t-1)
< s = \abs{\beta_0} ,
\]
where the third inequality follows from $D = \floor{(s-1)/(t-1)}$.
It follows that
\[
\beta_0 \setminus \parenv*{\cup_{j=1}^D \beta_j} \ne \varnothing ,
\]
namely, there is a point $x \in X$ such that $H_{x,\beta_0} = 1$
whereas $H_{x,\beta_j} = 0$ for all $1 \le j \le D$.
Hence, $H$ is $D$-disjunct.

Next, we consider the row weight, $\weight(H_x)$,
where $x$ is any point in~$X$:
this weight equals the number of blocks in~$\Block$ which contain~$x$.
Since $x$ is contained
in exactly one block in each parallel class and there are
in total $\rho$~parallel classes, we get $\weight(H_x) = \rho$.
\end{proof}

A \emph{transversal design} $\TD(s,g)$ is a triple $(X,\Group,\Block)$,
where $X$~is a set of $sg$ points, $\Group$~is a partition of~$X$
into $s$~partition elements (\emph{groups}),
each of size~$g$, and~$\Block$ is
a collection of $s$-subsets (\emph{blocks}) of~$X$ such that every
$2$-subset of~$X$ is contained either in one group or in one block,
but not both. A $\TD(s,g)$ is called \emph{resolvable} if its blocks
can be partitioned into parallel classes.

It is easy to see that in a $\TD(s,g)$, each block intersects with each
group at exactly one point. A direct calculation shows that
there are $g^2$ blocks and each point is contained in $g$~blocks.
So, if it is resolvable, then the blocks should be partitioned
into $g$~parallel classes.

It is known that the existence of a resolvable $\TD(s,g)$ is equivalent to the existence of $s-1$ mutually orthogonal latin squares of side $g$, while the latter can be constructed by using linear polynomials  (e.g., see Theorem 3.18 and Construction~3.29 in \cite[Section~III.3]{CD2007}). In the following example, we use linear polynomials to construct resolvable $\TD$s directly.

\begin{example}
\label{ex:TD}
Let~$q$ be a prime power. We can construct a resolvable $\TD(q,q)$
as follows. Take $X = \F_q\times \F_q$,
$\Group = \set*{\set{y} \times \F_q}_{y \in \F_q}$,
and $\Block = \set*{\beta_{a,b}}_{(a,b) \in \F_q \times \F_q}$,
where
\[
\beta_{a,b} = \set*{(y,ay+b) \,:\, y\in \F_q} .
\]
For each $a \in \F_q$, let $\cP_a = \set*{\beta_{a,b}}_{b\in \F_q}$;
clearly, $\set*{\cP_a}_{a \in \F_q}$ is a partition of~$\Block$.

Each block $\beta_{a,b}$ has size~$q$, which equals
the number of groups,
and every $2$-subset of the form $\set{(y,z),(y,z')}$
(which is contained in one group) cannot be contained in any block.
For a $2$-subset $\set{(y,z),(y',z')}$ with $y \ne y'$,
the system of equations
\[
 \begin{cases}
 ay+b =z\\
 ay'+b=z'
 \end{cases}
\]
has a unique solution for $(a,b)$;
hence, there is a unique block in~$\Block$ which contains
that $2$-subset. Therefore, $(X,\Group,\Block)$ is
a transversal design. Moreover, for each $a \in \F_q$ and each point
$(y,z)\in X$, there is a unique $b \in \F_q$ such that $ay+b = z$.
Hence, each $\cP_a$ is a parallel class.\qed
\end{example}

\begin{lemma}
\label{lem:TDtopacking}
If there is a resolvable $\TD(s,g)$, then for every $2 \le s' \le s$
and $1 \le g'\le g$, there is a $2$-$(s'g,s',1)$ packing design
with $g'$~parallel classes.
\end{lemma}

\begin{proof}
   From a resolvable $\TD(s,g)$ we can form a resolvable
$\TD(s',g)$ by deleting $s-s'$ groups. This resolvable design
consists of $g$~parallel classes. We can take~$g'$ of them to form
a $2$-$(s'g,s',1)$ packing design.
\end{proof}

\begin{theorem}
\label{thm:TDdisj-1}
Let $n, \rho, D \in \Z^+$ be such that $n/\rho$ is a prime power
and $D+1 \le \rho \le \sqrt{n}$. There is an explicit construction
of a $(D,\rho)$-disjunct $r \times n$ matrix
with constant row weight~$\rho$, constant column weight $D+1$,
and (therefore) number of rows
\[
r = \frac{(D+1)n}{\rho},
\]
thereby attaining the bound in Theorem~\ref{thm:lowbnd}.
\end{theorem}

\begin{proof}
Since $n/\rho$ is a prime power, we can take a resolvable
$\TD(n/\rho,n/\rho)$ from Example~\ref{ex:TD}.
Then, according to Lemma~\ref{lem:TDtopacking}, for any $D,\rho$
such that $D+1 \le \rho \le n/\rho$, we can construct
a $2$-$((D+1)n/\rho,D+1,1)$ packing with $\rho$~parallel classes. Since
each parallel class consists of $n/\rho$ blocks, the total number of
blocks is~$n$. So, the incidence matrix~$H$ of this packing is of order
$r \times n$, where $r = (D+1)n/\rho$, and constant row weight~$\rho$.
Moreover, according to Proposition~\ref{prop:incmatpacking},
$H$~is $D$-disjunct.
\end{proof}

Combining Theorem~\ref{thm:TDdisj-1} with Theorem~\ref{thm:disjtoerrloc}
yields the following result.

\begin{corollary}
\label{cor:code-2}
Let $n, \rho, \hp \in \Z^+$ be such that $n/\rho$ is a prime power
and $\hp \le \rho \le \sqrt{n}$. There is an explicit construction
of a linear $[n,k{\ge}n{-}r]$ code~$\code$ over~$\R$ with
\[
r = \frac{\hp n}{\rho}
\quad \textrm{and} \quad
\Gamma_\hp(\code) \le 2\rho=\frac{2\hp n}{r}.
\]
\end{corollary}

We note that the $\TD(q,q)$ in Example~\ref{ex:TD} is equivalent
to a $[q,2]$ (extended) RS code over~$\F_q$:
each block $\beta_{a,b}$ in
the~$\TD$ corresponds to a codeword whose positions are indexed by
the elements of $y \in \F_q$. An element $(y,z)$ contained in
the block indicates that in the corresponding codeword there should be
a symbol~$z$ at the position which is indexed by~$y$.

In the Kautz--Singleton construction,
the columns of the disjunct matrix are the codewords
of the binary code that is obtained by concatenating
a RS outer code over~$\F_q$ with
the binary code which consists of the words
in $\set{0,1}^q$ of Hamming weight~$1$.
In light of this, it is not difficult to see that the incidence matrix
of the~$\TD$ in Example~\ref{ex:TD} is actually
the disjunct matrix from the Kautz--Singleton construction with
an RS outer code of dimension~$2$. Hence,
the disjunct matrices in Theorem~\ref{thm:TDdisj-1} can also be obtained
by carefully choosing the columns of the Kautz--Singleton disjunct
matrix which correspond to the selected parallel classes.

We have the following product construction of~$\TD$'s, which yields
more disjunct matrices. The proof of this construction is
straightforward and is therefore omitted.

\begin{proposition}
\label{prop:product}
Let $(X,\Group,\Block)$ be a resolvable $\TD(s,g)$ with
a group partition $\Group = \set{\gamma_i}_{i \in \Int{s}}$
and with parallel classes $\cP_j$, $j \in \Int{g}$,
and let $(X',\Group',\Block')$ be a resolvable $\TD(s,g')$ with
a group partition $\Group' = \set{\gamma_i'}_{i \in \Int{s}}$
and with parallel classes $\cP_j'$, $j \in \Int{g'}$.
For any two blocks $\beta \in \Block$ and $\beta' \in \Block'$,
denote
\[
\beta \otimes \beta'
\eqdef \set{(x_i,x_i') \,:\, i \in \Int{s}},
\]
where $x_i$ (respectively, $x_i'$) is
the unique element in $\beta \cap \gamma_i$
(respectively, $\beta' \cap \gamma_i'$), $i \in \Int{s}$.
Then the set of points
\[
\bigcup_{i \in \Int{s}} (\gamma_i \times \gamma_i') ,
\]
the group partition
$\set{\gamma_i \times \gamma_i' \,:\, i \in \Int{s}}$,
and the set of blocks
\[
\set{\beta \otimes \beta' \,:\,
(\beta,\beta') \in \Block \times \Block'} ,
\]
form a resolvable $\TD(s,g g')$
with $g g'$ parallel classes
\[
\cP_{j,j}
\eqdef \set{\beta \otimes \beta' \,:\, 
(\beta,\beta') \in \cP_j \times \cP_{j'}'} ,
\quad j,j' \in \Int{g} \times \Int{g'}.
\]
\end{proposition}

\begin{theorem}
\label{thm:TDdisj-2}
Let $n, \rho$ be positive integers such that $\rho \le \sqrt{n}$
and $\rho$~divides~$n$, and let
$p_1^{e_1} p_2^{e_2}\cdots$ be the prime factorization
of $n/\rho$. Let $p^e = \min_i\set{p_i^{e_i}}$.

(i)~For any positive integer $D < \min \set{p^e,\rho}$,
there is an explicit construction of a $D$-disjunct
$r \times n$ matrix with constant row weight~$\rho$ and constant column
weight $D+1$, where $r = (D+1)n/\rho$
(thereby attaining the bound of Theorem~\ref{thm:lowbnd}).

(ii)~For any positive integer~$\hp$ such that
$\hp \le \min \set{p^e,\rho}$, there is
a linear $[n,k{\ge}n{-}r]$ code~$\code$ over~$\R$ with
\[
r = \frac{\hp n}{\rho}
\quad \textrm{and} \quad
\Gamma_\hp(\code) \le 2\rho=\frac{2\hp n}{r}.
\]
\end{theorem}

\begin{proof}
Since $p^e =\min_i\set{p_i^{e_i}}$, there is a resolvable
$\TD(p^e,p_i^{e_i})$ for each~$i$. Using the product construction
recursively, we obtain a resolvable $\TD(p^e,n/\rho)$.
Then, according to Lemma~\ref{lem:TDtopacking}, for any $D,\rho$
such that $D < \min \set{\rho,p^e}$ and $\rho \le n/\rho$,
we can construct a $2$-$((D+1)n/\rho,D+1,1)$ packing
with $\rho$~parallel classes.
Parts~(i) and~(ii) then follow from
Proposition~\ref{prop:incmatpacking} and
Theorem~\ref{thm:disjtoerrloc}, respectively.
\end{proof}


\appendix

\begin{proof}[Proof of Lemma~\ref{lem:RIP-alt}]
We first observe that the entries along the main diagonal of
$H^\transpose H$ are all~$1$
and that the absolute value of each off-diagonal entry is
at most~$\mu$. Hence,
\begin{eqnarray}
\norm{H \blde^\transpose}_2^2
& = &
\blde H^\transpose H \blde^\transpose \nonumber \\
& \ge &
\norm{\blde}_2^2 - \mu \sum_{0 \le i \ne j < n} \abs{e_i e_j}
\nonumber \\
& = &
(1 + \mu) \norm{\blde}_2^2
- \mu \sum_{i \in \Int{n}} \abs{e_i} \sum_{j \in \Int{n}} \abs{e_j}
\nonumber \\
\label{eq:objective1}
& = &
(1 + \mu) \norm{\blde}_2^2 - \mu \norm{\blde}_1^2 .
\end{eqnarray}

We next minimize the expression~\eqref{eq:objective1}
over~$\blde$ under the constraint that
$\blde \in \Ball(n,\hp)$ and~$\norm{\blde}_\infty$ is given.
Assuming without loss of generality that
$e_0 = \norm{\blde}_\infty$
and that $e_j = 0$ for all $j \in \Int{\hp:n}$,
we claim that the minimum is attained when
$\abs{e_1} = \abs{e_2} = \ldots = \abs{e_{\hp-1}}$. Otherwise, if
$\abs{e_i} \ne \abs{e_j}$ for some $1 \le i < j < \hp$ then
replacing both~$e_i$ and~$e_j$ by $(\abs{e_i} + \abs{e_j})/2$
would reduce the term $\norm{\blde}_2^2$ while
keeping $\norm{\blde}_1^2$ unchanged.

Substituting $\abs{e_i} \leftarrow x$ for all $i \in \Int{1:\hp}$
in~\eqref{eq:objective1} yields the following
quadratic expression in~$x$:
\begin{equation}
\label{eq:objective2}
(1 + \mu) (e_0^2 + (\hp-1) x^2) -
\mu (e_0 + (\hp-1) x)^2 .
\end{equation}
The coefficient of~$x^2$ is $(1 - (\hp-2) \mu)(\hp-1)$,
which is positive under our assumption
$\hp \le \ceil{1/\mu}$;
hence, \eqref{eq:objective2} attains a global minimum at
$x_{\min} = e_0 \cdot \eta_\hp(\mu)$.
Plugging this value into~\eqref{eq:objective2} yields the result.
\end{proof}

The lower bound in Lemma~\ref{lem:RIP-alt}
can be written more explicitly as
\[
\norm{H \blde^\transpose}_2^2
\ge
\frac{1 + \mu}{1 - (\hp-2) \mu}
\cdot \parenv*{1 - (\hp-1) \mu} \cdot \norm{\blde}_\infty^2 .
\]
Comparing with~\eqref{eq:cohtoRIP}, the bound in Lemma~\ref{lem:RIP-alt}
is expressed in terms of $\norm{\blde}_\infty$ rather than
$\norm{\blde}_2$, yet the multiplying constant therein is larger
when $\mu > 0$ and $\hp > 1$.

\bibliographystyle{IEEEtranS}
\bibliography{allbib}

\begin{thebibliography}{10}
\providecommand{\url}[1]{#1}
\csname url@samestyle\endcsname
\providecommand{\newblock}{\relax}
\providecommand{\bibinfo}[2]{#2}
\providecommand{\BIBentrySTDinterwordspacing}{\spaceskip=0pt\relax}
\providecommand{\BIBentryALTinterwordstretchfactor}{4}
\providecommand{\BIBentryALTinterwordspacing}{\spaceskip=\fontdimen2\font plus
\BIBentryALTinterwordstretchfactor\fontdimen3\font minus
  \fontdimen4\font\relax}
\providecommand{\BIBforeignlanguage}[2]{{%
\expandafter\ifx\csname l@#1\endcsname\relax
\typeout{** WARNING: IEEEtranS.bst: No hyphenation pattern has been}%
\typeout{** loaded for the language `#1'. Using the pattern for}%
\typeout{** the default language instead.}%
\else
\language=\csname l@#1\endcsname
\fi
#2}}
\providecommand{\BIBdecl}{\relax}
\BIBdecl

\bibitem{Boseretal91}
B.~E. Boser, E.~Sackinger, J.~Bromley, Y.~L. Cun, and L.~D. Jackel, ``An analog
  neural network processor with programmable topology,'' \emph{IEEE J.
  Solid-State Circuits}, vol.~26, no.~12, pp. 2017--2025, Dec. 1991.

\bibitem{Bouretal}
J.~Bourgain, S.~J. Dilworth, K.~Ford, S.~Konyagin, and D.~Kutzarova, ``Explicit
  constructions of {RIP} matrices and related problems,'' \emph{Duke Math. J.},
  vol. 159, no.~1, pp. 145--185, 2011.

\bibitem{Candes2008}
E.~Cand{\`e}s, ``The restricted isometry property and its implications for
  compressed sensing,'' \emph{C.R. Acad.\ Sci.\ Paris, Ser.~I}, vol. 346, pp.
  589--592, 2008.

\bibitem{CanRomTao06}
E.~Cand{\`e}s, J.~Romberg, and T.~Tao, ``Robust uncertainty principles: Exact
  signal reconstruction from highly incomplete frequency information,''
  \emph{IEEE Trans.\ Inf.\ Theory}, vol.~52, no.~2, pp. 489--509, Feb. 2006.

\bibitem{CD2007}
C.~Colbourne and J.~Dinitz, \emph{Handbook of combinatorial designs, second
  edition}.\hskip 1em plus 0.5em minus 0.4em\relax CRC press Boca Raton, FL,
  2007.

\bibitem{Donoho}
D.~Donoho, ``Compressed sensing,'' \emph{IEEE Trans.\ Inf.\ Theory}, vol.~52,
  no.~4, pp. 1289--1306, Apr. 2006.

\bibitem{DyaRyk82}
A.~G. D'yachkov and V.~V. Rykov, ``Bounds on the length of disjunctive codes,''
  \emph{Probl. Peredachi Inf.}, vol.~18, no.~3, pp. 7--13, 1982.

\bibitem{Fur96}
Z.~F{\"u}redi, ``On $r$-cover-free families,'' \emph{J.~Combin.~Theory Ser.~A},
  vol.~73, no.~1, pp. 172--173, 1996.

\bibitem{Huetal18}
M.~Hu, C.~E. Graves, C.~Li, Y.~Li, N.~Ge, E.~Montgomery, N.~Davila, H.~Jiang,
  R.~S. Williams, J.~J. Yang, Q.~Xia, and J.~P. Strachan, ``Memristor-based
  analog computation and neural network classification with a dot product
  engine,'' in \emph{Adv.\ Mater.}, vol.~30, Mar. 2018, paper no. 1705914.

\bibitem{Huetal16}
M.~Hu, J.~P. Strachan, Z.~Li, E.~M. Grafals, N.~Davila, C.~Graves, S.~Lam,
  N.~Ge, J.~Yang, and R.~S. Williams, ``Dot-product engine for neuromorphic
  computing: Programming {1T1M} crossbar to accelerate matrix-vector
  multiplication,'' in \emph{Proc. 53rd ACM/EDAC/IEEE Design Automat.\ Conf.\
  (DAC)}, Austin, TX, 2016, paper no. 19.

\bibitem{InaKaiOzg20}
H.~A. Inan, P.~Kairouz, and A.~{\"O}zg{\"u}r, ``Sparse combinatorial group
  testing,'' \emph{IEEE Trans.\ Inf.\ Theory}, vol.~66, no.~5, pp. 2729--2742,
  May 2020.

\bibitem{KatonaSeress93}
G.~Katona and A.~Seress, ``Greedy construction of nearly regular graphs,''
  \emph{European J.~of Combin.}, vol.~14, pp. 213--229, 1993.

\bibitem{KauSin64}
W.~Kautz and R.~Singleton, ``Nonrandom binary superimposed codes,'' \emph{IEEE
  Trans.\ Inf.\ Theory}, vol.~10, no.~4, pp. 363--377, Oct. 1964.

\bibitem{Kubetal90}
F.~J. Kub, K.~K. Moon, I.~A. Mack, and F.~M. Long, ``Programmable analog
  vector-matrix multipliers,'' \emph{IEEE J. Solid-State Circuits}, vol.~25,
  no.~1, pp. 207--214, Feb. 1990.

\bibitem{Lubell66}
D.~Lubell, ``A short proof of {S}perner's lemma,'' \emph{J.~Combin.~Theory
  Ser.~A}, vol.~1, no.~2, p. 299, 1966.

\bibitem{MacSlo78}
F.~J. MacWilliams and N.~J.~A. Sloane, \emph{The Theory of Error-Correcting
  Codes}.\hskip 1em plus 0.5em minus 0.4em\relax Amsterdam: North-Holland,
  1977.

\bibitem{PorRot11}
E.~Porat and A.~Rothschild, ``Explicit non-adaptive combinatorial group testing
  schemes,'' \emph{IEEE Trans.\ Inf.\ Theory}, vol.~57, no.~12, pp. 7982--7989,
  Dec. 2011.

\bibitem{Roth19}
R.~M. Roth, ``Fault-tolerant dot-product engines,'' \emph{IEEE Trans.\ Inf.\
  Theory}, vol.~65, no.~4, pp. 2046--2057, Apr. 2019.

\bibitem{Roth20}
------, ``Analog error-correcting codes,'' \emph{IEEE Trans.\ Inf.\ Theory},
  vol.~66, no.~7, pp. 4075--4088, Jul. 2020.

\bibitem{Roth2022ITW}
------, ``Fault-tolerant neuromorphic computing on nanoscale crossbar
  architectures,'' in \emph{Proc.\ 2020 IEEE Inf.\ Theory Workshop (ITW)},
  Mumbai, India, 2022, pp. 202--207.

\bibitem{Roth23}
------, ``Correction to ``analog error-correcting codes'','' \emph{IEEE Trans.\
  Inf.\ Theory}, vol.~69, no.~6, pp. 3793--3794, Jan. 2023.

\bibitem{Shafetal16}
A.~Shafiee, A.~Nag, N.~Muralimanohar, R.~Balasubramonian, J.~P. Strachan,
  M.~Hu, R.~S. Williams, and V.~Srikumar, ``{ISAAC}: A convolutional neural
  network accelerator with in-situ analog arithmetic in crossbars,'' in
  \emph{Proc.\ ACM/IEEE 43rd Annu.\ Int.\ Symp.\ Comput.\ Archit.\ (ISCA)},
  Seoul, Korea, Jun. 2016, pp. 14--26.

\end{thebibliography}

\end{document}